\documentclass[%
 aip, 
 pof,
 sd,%
 amsmath,amssymb,
preprint,%
]{revtex4-1}

\usepackage{graphicx}
\usepackage{dcolumn}
\usepackage{bm}

\newcommand{\eqr}[1]{(\ref{eq:#1})}

\newcommand{\etal}{\textit{et al.}}
\newcommand{\ie}{\textit{i.e.}}
\newcommand{\bs}[1]{\boldsymbol{#1}}

\begin{document}

\title{Droplet migration in a Hele--Shaw cell: Effect of the lubrication film on the droplet dynamics}

\author{Yue Ling}
\author{Jose-Maria Fullana}
\author{St\'ephane Popinet}
\author{Christophe Josserand}

\affiliation{ 
Institut Jean Le Rond d'Alembert, Sorbonne Universit\'es, UPMC Univ Paris 06, CNRS, UMR 7190, Paris, F-75005, France}%

\begin{abstract}
Droplet migration in a Hele--Shaw cell is a fundamental multiphase flow problem 
which is crucial for many microfluidics applications. We focus on the regime at low capillary number
and three-dimensional direct numerical simulations are performed to investigate the problem.  
In order to reduce the computational cost, an adaptive mesh is employed and high mesh resolution is only used 
near the interface. 
Parametric studies are performed on the droplet horizontal radius and the capillary number.
For droplets with an horizontal radius larger than half the channel height
the droplet overfills the channel and exhibits a pancake shape. 
A lubrication film is formed between the droplet and the wall and 
particular attention is paid to the effect of the lubrication film on the droplet velocity.  
The computed velocity of the pancake droplet is shown to be lower than the average inflow velocity, 
which is in agreement with experimental measurements. 
The numerical results show that both the strong shear induced by the lubrication film 
and the three-dimensional flow structure contribute to the low mobility of the droplet. 
In this low-migration-velocity scenario the interfacial flow in the droplet reference frame moves toward the rear 
on the top and reverses direction moving to the front from the two side edges. 
The velocity of the pancake droplet and the thickness of the lubrication film are observed to decrease 
with capillary number. 
The droplet velocity and its dependence on capillary number cannot be captured by the classic
Hele--Shaw equations, since the depth-averaged approximation neglects the effect of the lubrication
film. 
\end{abstract}

\maketitle

\section{INTRODUCTION}
\label{sec:intro}
Droplet-based microfluidics is a promising tool for performing biochemical or chemical assays 
or to observe liquid-liquid extraction \cite{Stone_2004a, Fair_2007a, Teh_2008a}. 
Droplets are unit systems of controlled volume and content, 
within which mixing, reacting and/or transferring  can be achieved. 
Therefore, a comprehensive understanding of droplet migration 
in microchannels is essential to many microfluidics applications. 
Rectangular channels with a width significantly larger than the height 
are commonly used in microfluidics devices. 
In such cases, the confinement of the droplet is only in the vertical direction
and the droplet is exposed to a two-dimensional Poiseuille flow (as in a Hele--Shaw cell). 
The migration of the droplet is driven by the ambient fluid 
with an average inflow velocity $U_f$ and a constant terminal migration velocity $U_d$ 
will be reached. 

\begin{figure}[tbpp]
	\centering
	\includegraphics[width=0.6\columnwidth]{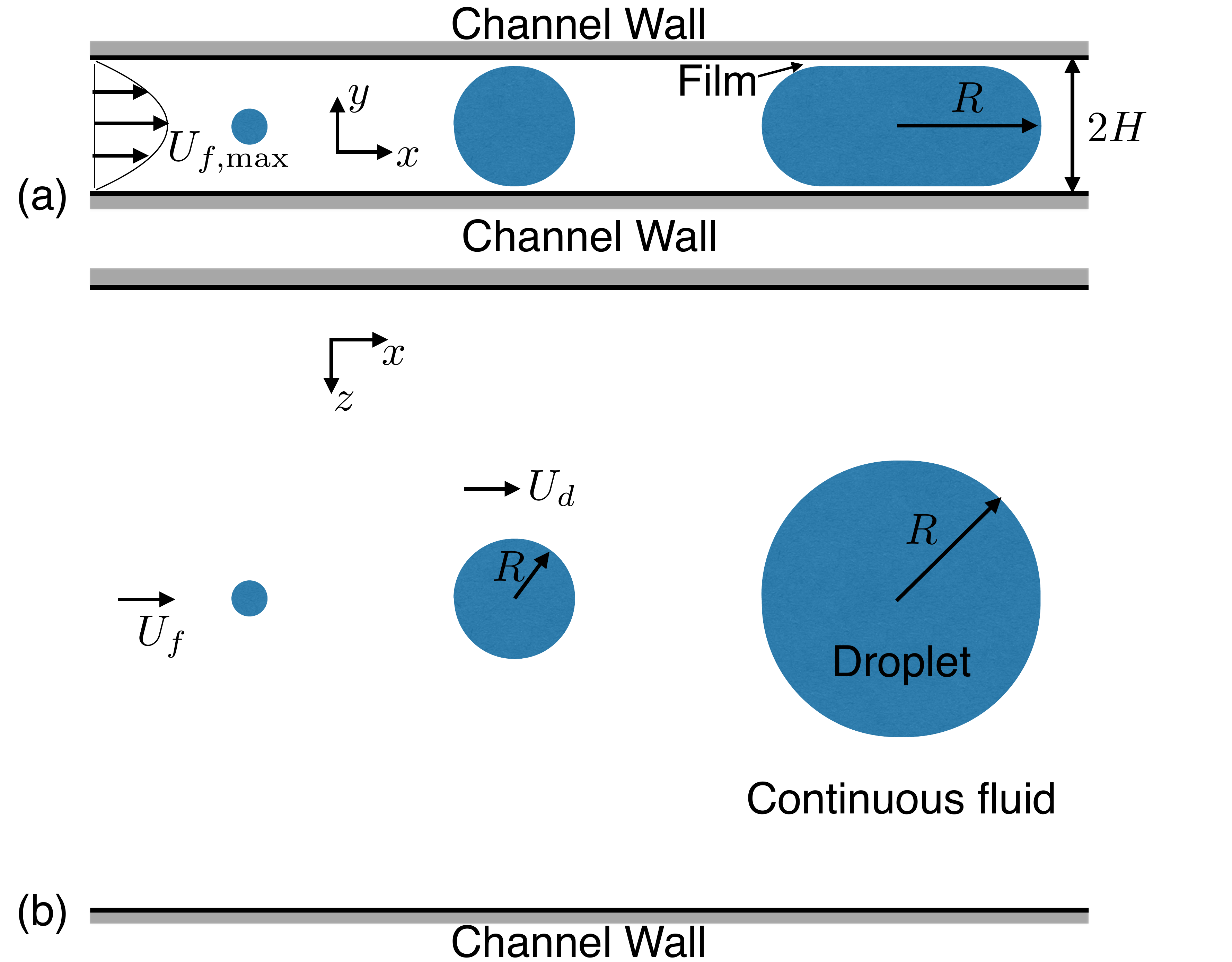} 
	\caption{Schematics of droplet migration in a Hele--Shaw cell for different droplet radii. 
	(a) Side-view and (b) top-view. }
	\label{fig:problem_description}
\end{figure}

The ratio between the horizontal radius of the droplet ($R$) 
and half of the channel height ($H$) has been shown to have a strong effect on 
the droplet migration velocity $U_d$ in a microchannel (hereafter we simply refer to
$U_d$ as the droplet velocity since the transient process is not of interest in the present study).
The overall variation of $U_d$ with $R/H$ can be seen in the experiments by Roberts \etal \cite{Roberts_2014a},
(the experimental results will be shown later in section \ref{sec:3D})
and can be divided into three different regimes according to the dominant mechanism: 
\begin{itemize}
	\item Poiseuille-dominated regime (Regime P), $R/H\ll 1$. 
	In this regime the droplet velocity is dominated by the Poiseuille flow in the Hele--Shaw cell. 
	In the limit of $R/H\to 0$  (as the droplet is a point mass)  
	the droplet at the centerline of the channel will move with 
	the maximum fluid velocity ($U_{f,max}=1.5 U_f$).  
	As $R$ increases the droplet is exposed to the parabolic fluid velocity profile 
	with a decreasing velocity toward the top and bottom walls and the droplet slows down. 
	Theoretical studies on the migration of droplets much smaller than the channel height 
	have been conducted in many previous works, see for example Nadim and Stone \cite{Nadim_1991a} 
	and Hudson \cite{Hudson_2010a}. 
	
	\item Film-dominated regime (Regime F), $R/H\gg 1$: In this regime, 
	the droplet overfills the channel and a thin lubrication film is formed between 
	the droplet and the wall. The droplet loses its roughly spherical geometry 
	and becomes ``pancake-like''. Furthermore, the droplet velocity $U_d$ becomes independent of $R/H$
	and is mainly dictated by the lubrication film.
	The film thickness $h$ is governed by the balance between surface tension and viscous stresses.
	As the former tends to minimize the interface curvature and pushes the droplet interface 
	closer to the wall, the latter tends to open up the gap between the droplet and the wall. 
	Therefore the capillary number ($\mathrm{Ca}=\mu_f U_d/\sigma$, where $\mu_f$ and $\sigma$ 
	is the viscosity of the ambient fluid 
	and the surface tension, respectively), which is the ratio between the viscous and surface tension forces, 
	is critical in determining the film thickness and the droplet mobility in this regime \cite{Bretherton_1961a}. 
	Beyond the capillary number, the viscosity ratio between the droplet 
	and the ambient fluid has also been shown to have a significant influence on
	the lubrication film thickness and the flow pattern \cite{Hodges_2004a}. 

	\item Transition regime (Regime T), $R/H \approx 1$: This regime connects the previous two regimes. 
	As the droplet radius is comparable to half of the channel height, the effect of confinement on
	the droplet becomes significant. As a result, the droplet velocity decreases dramatically with $R/H$ 
	in this regime. 
\end{itemize}
The three droplets shown in Fig.\ \ref{fig:problem_description} schematically represent these three regimes.
The axes corresponding to the channel length, height, and width are defined as 
$x$, $y$, and $z$, respectively.
The experimental and numerical results of Roberts \etal \cite{Roberts_2014a} clearly show the variation 
of droplet velocity in the Poiseuille-dominated regime, and close agreement with theoretical predictions 
has been achieved. The experimental data in the film-dominated regime is less documented however. Furthermore, simulation results for the transition and film-dominated regimes are not provided,
due to the numerical challenge in resolving the thin film \cite{Roberts_2014a}. The present study aims at improving 
the current understanding of droplet motion in microchannels, in particular the droplet dynamics in the 
transition and film-dominated regimes. 

In contrast with scenarios in which the droplet is much smaller than the channel height, 
the dynamics of droplets in the film-dominated regime (\ie, $R/H\gg1$) 
are more complicated and less understood. 
Early efforts in predicting the velocity of a flat bubble in a Hele--Shaw cell 
go back to the pioneering theoretical work by Taylor and Saffman \cite{Taylor_1959a}.
The Hele--Shaw equations are obtained by averaging the Stokes equations along the channel height (depth),
and it can be shown that the bubble velocity $U_d$ is always larger than the average inflow velocity $U_f$.
For low capillary numbers, the horizontal cross-section of the bubble is close to a circle 
(exhibiting a pancake shape similar to the present problem) \cite{Kopf-Sill_1988a}, 
then $U_d/U_f=2$ according to the Taylor--Saffman theory \cite{Taylor_1959a, Tanveer_1986a, Tanveer_1987a}. In the later experiments by Kopf-Sill and Homsy \cite{Kopf-Sill_1988a}, 
the bubble velocity was observed to be about an order of magnitude lower than the Taylor--Saffman prediction. 
Possible reasons for the discrepancy have been explored, such as an excessive drag due to a moving contact line 
within the film and Marangoni effects due to surfactant transport 
\cite{Saffman_1989a, Park_1994a, Maruvada_1996a}, yet the underlying mechanisms responsible 
for the low mobility of the bubble is still not fully understood. In the limit when the 2D droplet is perfectly circular, 
the Hele--Shaw equations have also been solved analytically by Afkhami and Renardy \cite{Afkhami_2013a} and 
Gallaire \etal \cite{Gallaire_2014a} to obtain the droplet velocity for arbitrary viscosity ratios as 
\begin{equation}
	\frac{U_d}{U_f}= \frac{2}{1+\mu_d/\mu_f}\, , \label{eq:HS_drop_vel}
\end{equation}
where $\mu_d$ is the viscosity of the droplet fluid. Equation \eqr{HS_drop_vel} recovers
the Taylor--Saffman solution $U_d/U_f=2$ for bubbles ($\mu_d/\mu_f=0$). 
Nevertheless, since the Hele--Shaw equations ignore the lubrication film, 
Eq.\ \eqr{HS_drop_vel} does not capture the effect of the film dynamics 
on the droplet velocity. 

Due to the small scale of microchannels, 
experimental measurements of the detailed flow structure are very difficult. 
Recently Huerre  \etal \cite{Huerre_2015a} presented quantitative optical measurements 
of lubrication film thicknesses as small as 20 nm for a channel height $2H=25 \mu$m (For such a thin film, the intermolecular forces become important for the film dynamics 
and droplet velocity. In the present study, we focus on the cases where
the minimum film thickness is always larger than one micrometer. As a result,
the effects of intermolecular forces can be ignored.)
However it is very challenging to probe local values of velocity and pressure, 
and consequently the flow structure inside and outside of the droplet is hard to obtain. 
Numerical simulations which provide a complete description 
of the flow field can shed light on the present problem. 
It has been shown that numerical approximations of the Navier-Stokes equations 
can represent flow physics in many microfluidics applications, 
such as droplets/bubbles migration and the breakup of droplets at T-junctions 
\cite{Afkhami_2011a, Bedram_2013a, Hoang_2013a, Hoang_2013b}.

Nevertheless 3D direct numerical simulations of droplet migration in a microchannel are challenging as well. 
This is in particular true for droplets with large ratio $R/H$ and small Ca. 
Since capillary effects are dominant, the time step used in the temporal integration 
is controlled by the surface tension term in the momentum equation. 
For small capillary numbers the time step becomes very small 
and thus the number of steps for the terminal velocity of droplet to be reached is huge. 
In addition, at low capillary numbers the thickness of the lubrication film becomes very small 
(several orders of magnitude smaller than the channel height) \cite{Huerre_2015a}. 
Since the mesh resolution and the time step restriction 
are dictated by the smallest length scale in the problem, 
the total number of cells and integration steps in 3D simulations are both very large. 
Three-dimensional simulation results of droplet/bubble in microchannels 
are rarely seen until the very recent works by Hoang \etal \cite{Hoang_2013b}, 
Roberts \etal \cite{Roberts_2014a} and Zhu and Gallaire \cite{Zhu_2016a}. 
However, as far as the authors are aware, a 3D direct numerical solution to the Navier-Stokes equations 
for droplets with a large $R/H$ and with a lubrication film formed has never been reported
in the literature. 

The goal of the present study is to investigate the droplet migration in a Hele--Shaw cell 
through 3D direct numerical simulations. 
The capillary number of the droplet is considered to be small enough so that 
the horizontal shape of the droplet is near-circular all the time. 
Particular attention will be paid to the effect of the lubrication film 
on the droplet dynamics. 
A two-phase flow solver, Gerris\cite{Popinet_2003a, Popinet_2009a}, is employed 
to solve the variable-density Navier-Stokes equation with surface tension. 
In the Gerris solver, the Volume-of-Fluid (VOF) method is used to resolve the interface 
between the two phases and an adaptive mesh is used to refine the mesh only in the lubrication film 
and near the interface, so that the computational cost can be reduced. 
The governing equations and the numerical methods are presented in section \ref{sec:equations}. 
The simulation results will be shown in sections \ref{sec:2D} and \ref{sec:3D}. 
We will first perform simulations of 2D droplets migrating in a microchannel 
with parameters similar to the 3D case for the purpose of validation, see \ref{sec:2D}. 
The simulation results and discussions for the 3D droplet are then presented in section \ref{sec:3D}. 
Finally, we draw conclusions on the present study. 

\section{Governing equations and numerical methods}
\label{sec:equations}

\subsection{Governing equations}
\label{sec:gov_eqns}
The one-fluid approach is employed to resolve the two-phase flow, 
where the phases corresponding to the droplet and the ambient fluid 
are treated as one fluid with material properties that change abruptly across the interface. 
The incompressible, variable-density, Navier-Stokes equations 
with surface tension can be written as
\begin{align}
	\rho (\partial_t\bs{u} + \bs{u} \cdot \nabla \bs{u}) = -\nabla p 
	+ \nabla \cdot (2\mu \bs{D}) + \sigma \kappa \delta_s \bs{n}\, , 
	\label{eq:mom} \\
	\nabla \cdot \bs{u} = 0 \, ,
	\label{eq:divfree}
\end{align}
where $\rho$, $\mu$, $\bs{u}$, and $p$ represent density and viscosity, velocity and pressure, respectively. 
The deformation tensor is denoted by $\bs{D}$ with components $D_{ij}=(\partial_i u_j + \partial_j u_i)/2$. 
The third term on the right hand side of Eq.\ \eqr{mom} is a singular term, with 
a Dirac distribution function $\delta_s$ localized on the interface, and it represents 
the surface tension force. The surface tension coefficient is $\sigma$, 
and $\kappa$ and $\bs{n}$ are the local curvature and unit normal of the interface. 

The volume fraction $C$ is introduced to distinguish the different phases, 
in particular $C=1$ in the computational cells with only the ambient fluid (respectively $C=0$ in
the droplet), 
and its time evolution satisfies the advection equation
\begin{align}
	\partial_t C + \bs{u} \cdot \nabla C =0 \, .
	\label{eq:color_func}
\end{align}
The fluid density and viscosity are then defined by
\begin{align}
	\rho  & = C \rho_f + (1-C) \rho_d \, , 
	\label{eq:density} \\
	\mu  & = C \mu_f + (1-C) \mu_d \, .
	\label{eq:viscosity}
\end{align}
where the subscripts $f$ and $d$ represent the ambient fluid and the droplet respectively. 

\subsection{Numerical methods}
\label{sec:numerics}
The Navier-Stokes equations (Eqs.\ \eqr{mom} and \eqr{divfree}) are solved by
the open-source package Gerris\cite{Popinet_2003a,Popinet_2009a}. 
In Gerris, a finite volume approach based on a projection method is employed.   
A staggered-in-time discretisation of the volume-fraction/density and pressure 
leads to a formally second-order accurate time discretisation. 
The interface between the different fluids are tracked and followed 
using a Volume-of-Fluid (VOF) method \cite{Scardovelli_1999a,Tryggvason_2011a}. 
A quad/octree spatial discretisation is used, which gives a very important flexibility 
allowing dynamic grid refinement into user-defined regions \cite{Popinet_2003a}. 
Finally the height-function (HF) method is utilized to calculate the local interface curvature, 
and a balanced-force surface tension discretization is used \cite{Francois_2006a, Popinet_2009a}. 

\subsection{Time step restriction}
\label{sec:numerics}
Both temporal and spatial schemes of integration are explicit. Therefore, 
numerical stability requires the time step for integration to satisfy the stability conditions 
corresponding to the advection, viscous, and surface tension terms, respectively. 
Due to the small capillary number in the present problem, the time step restriction 
is dominated by the contribution from the surface tension force. The conventional 
capillary time step restriction is given by Brackbill \etal \cite{Brackbill_1992a} as, 
\begin{align}
	\Delta t \le \Delta t_{BKZ} = \sqrt{ { \rho \Delta^3 \over \pi \sigma } }\, \label{eq:dt_surf_tension}\, ,
\end{align}
where $\Delta$ represents the cell size. For droplets with large $R/H$, 
the thickness of the lubrication film is the smallest scale in the problem 
and thus dictates the smallest cell size and the time step to be used.
Taking the conventional scaling of film thickness \cite{Landau_1942a, Bretherton_1961a}, 
the grid size can be estimated as  $\Delta \sim h\sim H Ca^{2/3}$. 
(The scaling of the film thickness on capillary number will be discussed in detail in the sections \ref{sec:2D} and \ref{sec:3D}.)  
Then Eq.\ \eqr{dt_surf_tension} becomes 
\begin{align}
	\Delta t_{BKZ} = \sqrt{ { \rho H^3 \mathrm{Ca}^2 \over \pi \sigma } }\, \label{eq:dt_surf_tension_BKZ}\, . 
\end{align}
If we assume that the terminal migration velocity is reached after a time of order $H/U_d$, then 
the total number of time steps can be estimated as 
\begin{align}
	\mathcal{N}_t \ge \sqrt{ {\pi \sigma  \over  \rho H \mathrm{Ca}^2 U_d^2} } \sim \mathrm{Ca}^{-3/2} \mathrm{Re}^{-1/2}\, \label{eq:Nt_surf_tension_BKZ}\, , 
\end{align}
where $\mathrm{Re}=\rho_f U_d H/ \mu_f$. 
As the total number of grid points (on a regular Cartesian grid) will scale like 
$\mathcal{N}_{\Delta} \sim (H/\Delta)^D \sim \mathrm{Ca}^{-2D/3}$ 
with $D$ the dimension of the problem. Then the total work (number of grid points times number of time steps)  for a 3D problem will thus scale as 
\begin{align}
	\mathcal{N}_{\Delta} \mathcal{N}_t \sim \mathrm{Ca}^{-7/2} Re^{-1/2}\, \label{eq:totalwork_surf_tension_BKZ}\, . 
\end{align}

Given that capillary numbers for microfluidics applications can be as low as $10^{-5}$, we see that direct numerical simulations will be extremely expensive. The adaptive mesh refinement in Gerris allows to refine the mesh only in the lubrification film, which greatly decreases this cost, however this is still not sufficient to reach such low capillary numbers.

A different capillary time step restriction has been proposed by Galusinski and Vigneaux \cite{Galusinski_2008a}.
They suggested that when both $\mathrm{Re}$ and $\mathrm{Ca}$ are very small, the capillary time step 
should satisfy 
\begin{align}
	\Delta t \le  \max(\Delta t_{BKZ}, \Delta t_{STK})\, \label{eq:dt_GV}\, ,
\end{align}
where $\Delta t_{STK} = \mu_f \Delta / \sigma$ is the capillary time step related to the Stokes equation
(when inertial effects are negligible).  It can then easily be shown that the ratio 
$\Delta t_{BKZ}/\Delta t_{STK}\sim \mathrm{Re}^{-1/2}\mathrm{Ca}^{-1/6}$. Therefore,  
if $\mathrm{Re}< \mathrm{Ca}^{1/3}$ the larger time step $\Delta t_{STK}$ can be used. 
In the present study, however, we are interested in the cases where 
$\mathrm{Re}\sim O(1)$ and $\mathrm{Ca}$ varies from $10^{-3}$ to $10^{-1}$ (see the detailed values in 
section \ref{sec:3D}). As a consequence, the criterion $\mathrm{Re}< \mathrm{Ca}^{1/3}$ is never satisfied 
and Eqs.\ \eqr{dt_surf_tension}-\eqr{totalwork_surf_tension_BKZ} remain valid to estimate the time step and computational cost.

\section{2D simulations of droplet migration in a microchannel}
\label{sec:2D}
\begin{figure}[tbp]
\centering
\includegraphics[width=0.8\columnwidth]{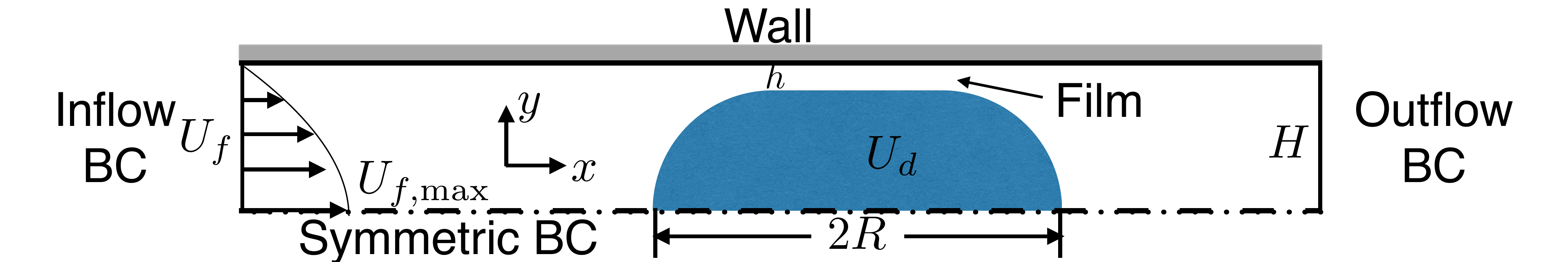}
\caption{Two-dimensional simulation setup for a droplet in a planar microchannel.}
\label{fig:problem_description_2d}
\end{figure}

Before we present the 3D results, 2D simulations for a droplet moving 
in a microchannel are performed for validation purpose. 
The simulation setup is schematically shown in Fig.\ \ref{fig:problem_description_2d}. 
The flow is symmetric in the $y$ direction with regard to the axis $y=0$, 
thus only the upper half of the channel is considered. 
The inflow from the left boundary is a parabolic velocity profile with mean value $U_f$ and maximum value $U_{f,max}=1.5 U_f$. 
An outflow boundary condition is applied at the right boundary of the domain; 
while the top boundary is taken as a no-slip wall condition. 
The length of the bubble is denoted by $2R$, and $R/H=2$. 
As the droplet is constrained by the channel, a lubrication film is formed 
between the droplet and the wall. 
The channel length is $L_x=16H$, which is long enough for the droplet 
to reach its terminal velocity $U_d$ before being affected by the outflow boundary. 

The physical properties of the ambient fluid and the droplet are listed in 
Table \ref{tab:phys_parameters}, which are similar to the recent experiments by 
Roberts \etal \cite{Roberts_2014a}
The Reynolds number based on the channel height and average inflow velocity, 
$\mathrm{Re}_f=\rho_f U_f H/\mu_f$, is taken to be 1.35. 
The surface tension is varied from 0.0002 to 0.0159 N/m. 

\begin{table}[tbp]
\caption{\label{tab:phys_parameters} Physical properties of the ambient fluid and the droplet in the present study. \footnote{Symbol 
with $\hat{}$ denotes dimensional parameters.} }
\begin{ruledtabular}
\begin{tabular}{cccccc}
$\hat{\rho}_f$ (kg/m$^3$)	& $\hat{\rho}_d$ (kg/m$^3$) & $\hat{\mu}_f$ (Pa\ s)	& $\hat{\mu}_d$ (Pa\ s) &	$\hat{H}$ ($\mu$m) & $\hat{U}_f$ (m/s) \\
\hline
750	& 1000 & $7.4\times 10^{-4}$ & $ 10^{-3}$ & $50$ & 0.0267\\
\end{tabular}
\end{ruledtabular}
\end{table}

Bretherton \cite{Bretherton_1961a} studied theoretically and experimentally 
the dynamics of a long bubble in a cylindrical channel. 
In the limit of small capillary numbers, \ie, $Ca=\mu_f U_d/\sigma \ll 1$, 
it is shown that the ratio between the film thickness $h$ 
and the half-height of the channel $H$ scales as $Ca^{2/3}$, as does the ratio between the excess of 
droplet velocity compared to the mean inflow velocity of the ambient fluid:
\begin{eqnarray}
{ h \over H } = { (U_d-U_f) \over U_f} \sim \mathrm{Ca}^{2/3}
\end{eqnarray}
the proportionality constant depends on the geometrical configuration (planar or cylindrical channels) 
and the viscosity ratio, although temperature and surfactants may also modify the surface tension.
It can be shown that $h/H$ is identical to $(U_d-U_f)/U_f$ due to mass fluxes balance.  

In the original work of Bretherton \cite{Bretherton_1961a}, 
the viscosity of the fluid within the bubble is considered to be zero. 
An extension to droplets of arbitrary viscosity was given later by Hodges \etal \cite{Hodges_2004a}
The theory for a slender droplet in a cylindrical channel 
is readily modified for a 2D droplet confined by two parallel plates. 
The film thickness and the excess of droplet velocity for a 2D droplet 
in the limits of $\mu_d\to 0$ (bubbles) and $\mu_d\to \infty$ (very viscous droplets) are given by
\begin{equation}
	\frac{h}{H} = \frac{U_d - U_f }{U_f} \sim 0.643 \ (3 \mathrm{Ca})^{2/3}\, .
\label{eq:scaling_2d_inviscid}
\end{equation}
and 
\begin{equation}
	\frac{h}{H} = \frac{U_d - U_f }{U_f} \sim 0.51 \ (3 \mathrm{Ca})^{2/3}\, , 
\label{eq:scaling_2d_visc}
\end{equation}
respectively \cite{Afkhami_2011a}. 
Similar scaling relations have also been derived for moving foams as well as bubble trains in micro-devices \cite{Baroud_2010a}. 


The multi-scale nature of the problem can be seen through 
Eqs.\ \eqr{scaling_2d_inviscid} and \eqr{scaling_2d_visc}. 
For a capillary number of $10^{-4}$, $h/H=0.00288$ and therefore 
the length scales involved in the system span over three orders of magnitudes. 
This fact highlights the importance of the adaptive mesh refinement technique for the present study, 
which reduces the overall computational costs by refining the mesh only in the lubrication film 
and near the interface. 

\begin{figure}[tbp]
\centering
\includegraphics[width=0.99\columnwidth]{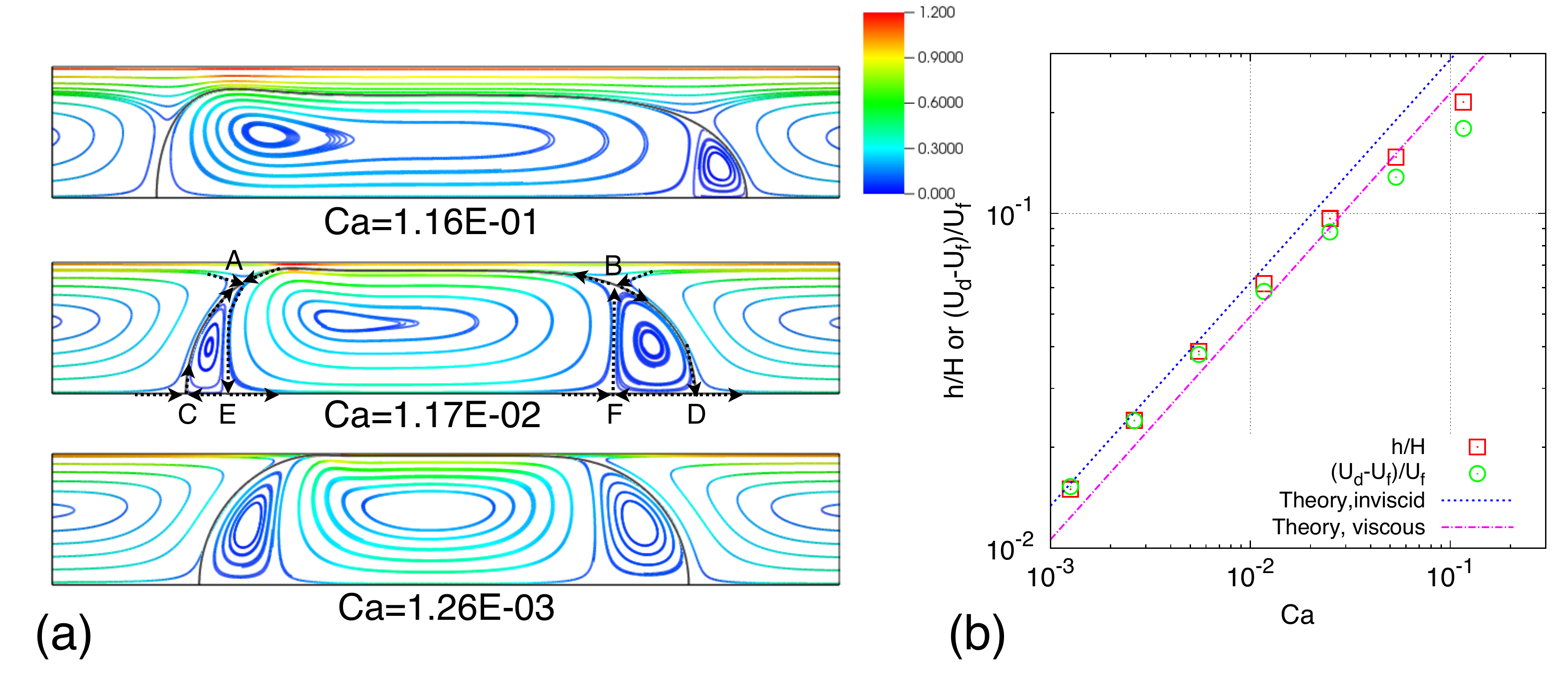}
\caption{(a) Droplet shape and the streamlines in the droplet reference frame. The color of the streamlines denotes the velocity magnitude in the droplet reference frame. (b) Relative droplet velocity and film thickness as functions of the capillary number. }
\label{fig:2d_results}
\end{figure}

The shapes of the droplets and the corresponding streamlines in the droplet reference frame 
are shown in Fig.\ \ref{fig:2d_results}(a) for different capillary numbers. 
The color of the streamlines denotes the velocity magnitude in the droplet reference frame. 
The droplet shape and flow pattern are in full agreement with the theory
\cite{Bretherton_1961a, Hodges_2004a}. Under the effect of the viscous stress, 
the radius of curvature at the front meniscus is smaller than that at the rear. A small bump is visible on the top-rear of the droplet.
The four stagnation points on the droplet interface (A: top-rear; B: top-front; C: rear; D: front) 
and the other two inside the droplet (E: near the rear; F: near the front) are clearly seen. 
For small Ca, the major recirculation (A-E-F-B-A) is located at the center of the droplet, 
with two minor ones sitting on both sides (A-E-C-A and B-F-D-B). 
When Ca increases the major recirculation shifts toward the rear of the droplet and the 
two minor recirculation zones become smaller. When the capillary number reaches 
$\mathrm{Ca}=0.116$ the stagnation points C and E get very close and  
the minor recirculation near the rear disappears. 

Figure \ref{fig:2d_results}(b) shows a quantitative comparison between the numerical 
and the theoretical results for the droplet velocity and the film thickness. 
The excess of droplet velocity $(U_d - U_f)/U_f$ (which is the same as $h/H$) are plotted 
as functions of the droplet capillary number $\mathrm{Ca}$ in log-log scale, 
for $\mathrm{Ca}$ varying from $1.24 \times 10^{-3}$ to $9.87 \times 10^{-2}$. 
The droplet velocity in the simulation is evaluated by a weighted-integral over all the cells occupied 
by the droplet fluid, \ie, $ U_d =  \int_S u\ (1-C) \ dS / \int_S (1-C)\  dS$. 
It is seen that the simulation results match the theoretical results quite well. 
For large $\mathrm{Ca}$ the simulation results are slightly lower than the theoretical prediction. 
This observation is consistent with the previous studies by Afkhami \etal \cite{Afkhami_2011a}


\section{3D simulations of droplet migration in a microchannel (Hele--Shaw cell)}\label{sec:3D}
In this section we present the 3D simulation results of droplet migration in a Hele--Shaw cell. 
The physical properties of the ambient fluid and the droplet are identical to those of the previous 2D simulations, 
but the droplet is 3D and the cross-section in the $x$-$z$ plane is close to a circle. 

\subsection{Simulation setup}
\begin{figure}[tbp]
	\centering
	\includegraphics[width=0.9\columnwidth]{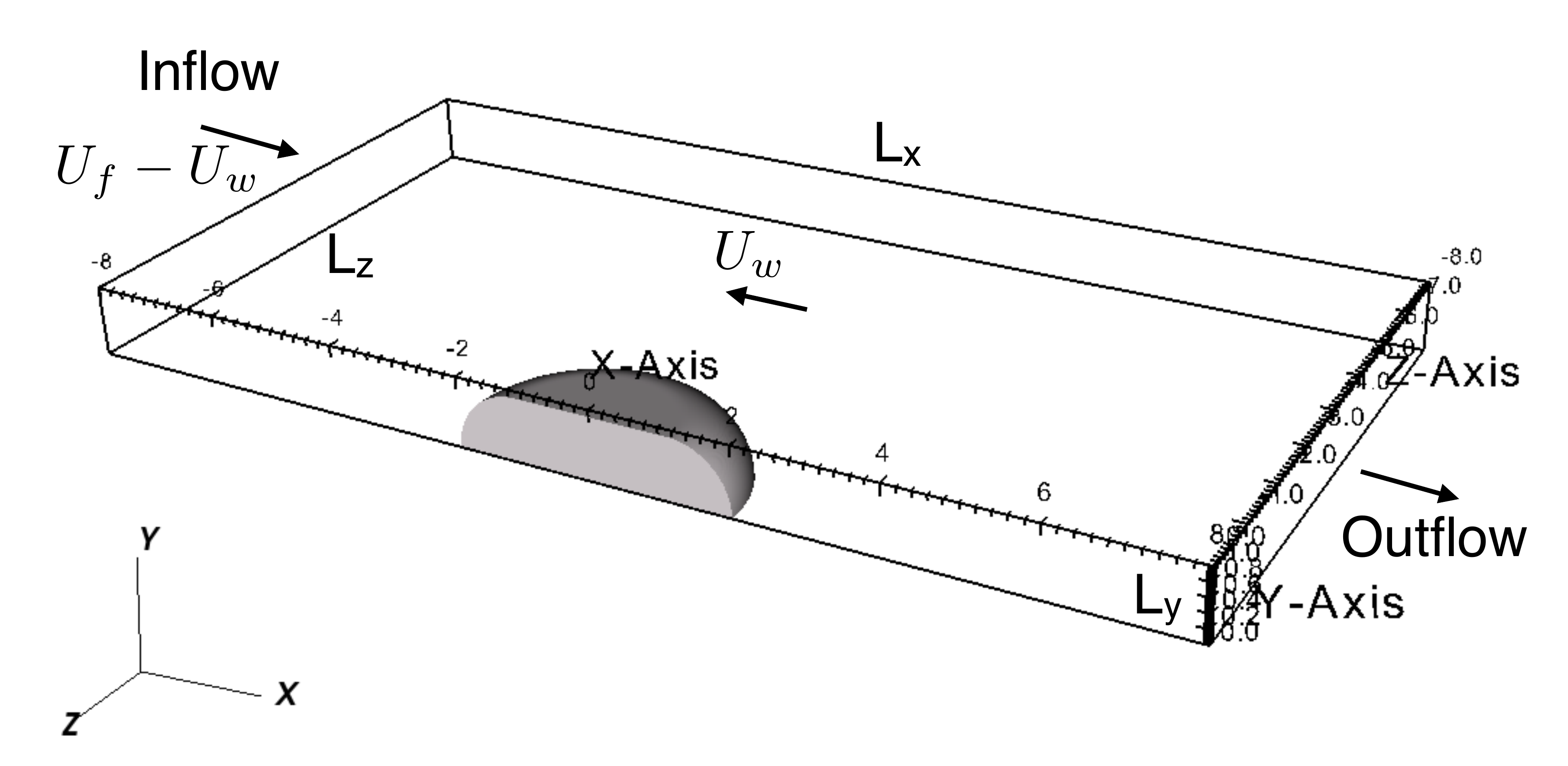} 
	\caption{Three-dimensional simulation setup for a droplet in a Hele--Shaw cell.}
	\label{fig:3d_setup}
\end{figure}

The computational domain for a 3D droplet migrating in a Hele--Shaw cell is shown in Fig.\ \ref{fig:3d_setup}. 
It is assumed that the droplet motion and the corresponding flow is symmetrical in the $y$ and $z$ directions, 
with respect to the planes $y=0$ and $z=0$, respectively. Therefore, only a quarter of the overall 
domain is actually computed. Taking the half-height of the channel $H$ as the typical length scale, 
the length, height, and width of the computational domain are taken as $L_x=16H$, $L_y=H$, 
and $L_z=8H$, respectively. 

An inflow boundary condition is imposed on the left of the domain with a parabolic velocity profile. 
Correspondingly an outflow boundary condition is used on the right. The top of the domain is 
taken to be a no-slip wall moving with a velocity $U_w$ in the opposite direction with respect to the flow. 
Therefore the simulation is indeed conducted in a moving reference frame with a mean 
inflow velocity of $U_f-U_w$. 
The purpose of using a moving reference frame is to keep the droplet near the center 
of the domain and to avoid the influence of the boundaries. Ideally the terminal droplet velocity 
in the lab reference frame $U_d$ should be used for $U_w$, but since $U_d$ is unknown \textit{a priori}, 
an initial guess of $U_d$ is used for $U_w$. Nevertheless as long as the droplet does not 
get too close to the inflow and outflow boundaries, the specific value of $U_w$ is immaterial 
to the simulation results. Finally the other boundaries are set to symmetric boundary conditions. 

The droplet is taken to be initially stationary in the simulation frame. 
Driven by the ambient fluid flow the droplet approaches the terminal droplet velocity 
as time evolves. The initial shape of droplets with $R/H < 1$  
is taken to be a sphere with radius $R$. For $R/H \gtrsim 1$ the cross-section at $y=0$ is 
a circle with radius $R_0$. The top of the droplet is initially flat with a constant distance of $0.1H$ 
to the top wall. The ``edge" of the droplet is formed by a quarter of a circle of radius of $0.9H$ rotated with respect 
to the $y$ axis, see Fig.\ \ref{fig:3d_setup}. Similar to the droplet velocity, the droplet will evolve 
to the final shape for which the surface tension and viscous forces at the interface are in balance. 
The final value of $R$, based on the horizontal cross-section area $S_{y=0}$
(at the plane $y=0$)  as $R=\sqrt{2S_{y=0}/\pi}$, is slightly different from its initial value $R_0$.
Due to the initial configuration we impose, the difference between $R$ and $R_0$ is generally small. 
(Therefore from here on we do not distinguish $R_0$ 
and $R$ unless otherwise indicated.)
Similar to the velocity of the top wall, the specific details of the initial shape of the droplet 
are  irrelevant to the final results as long as the terminal droplet velocity and shape are reached. 

An adaptive mesh is used to discretize the spatial domain. The finest cell size used is $\Delta_{\min}=H/128$. 
The mesh is refined at the interface based on the gradient of the volume fraction in each cell. 
Mesh refinement is also conducted using an \textit{a posteriori} error estimate of all three velocity components 
as a cost function for adaptation. 
A mesh refinement study was performed in the 2D simulations, showing that $\Delta_{\min}$ is sufficient 
for the range of capillary numbers considered in the present study. 


\begin{table}[tbp]
\caption{\label{tab:dmls_parameters} Key dimensionless parameters in 3D simulations.  }
\begin{ruledtabular}
\begin{tabular}{ccccc}
$r=\rho_f$/$\rho_d$  & $m=\mu_f/\mu_d$ & $\mathrm{Re}_f=\rho_f U_f H/\mu_f$ & $R/H$ & $\mathrm{Ca}_f=\mu_f U_f/\sigma$\\
\hline
0.75 & 0.74 & 1.35 & 0.5 to 2 & $6.17\times10^{-3}$ to $9.87\times10^{-2}$\\
\end{tabular}
\end{ruledtabular}
\end{table}

The physical parameters are identical to the 2D simulation presented in section \ref{sec:2D} 
(see Table \ref{tab:phys_parameters}). 
Parametric studies are performed by varying the surface tension coefficient $\sigma$ and the droplet radius $R$. 
Taking the ambient fluid density 
$\rho_f$, the half of the channel height $H$, and the mean inflow velocity $U_f$ as characteristic scales, 
the unknown variables and physical parameters can be non-dimensionalized 
and the governing equations (Eqs.\ \eqr{mom}--\eqr{color_func})
are solved in dimensionless form. The key dimensionless parameters of the present problem are
given in Table \ref{tab:dmls_parameters}. As shown in the table, the ratio $R/H$ and the capillary number 
based on the average inflow velocity $\mathrm{Ca}_f$ are chosen as the tuning parameters.  

The capillary numbers in the 3D simulation are significantly larger than that in experiments by 
Roberts \etal \cite{Roberts_2014a}, in which $\mathrm{Ca}_{f,\mathrm{Exp}}=9.87\times 10^{-4}$. 
The smallest $\mathrm{Ca}_f$ considered here is about an order of magnitude larger than this experimental value. 
As illustrated in the previous 2D results, the lubrication film formed is very small for small capillary numbers. 
It is prohibitively difficult to simulate droplets with a larger $R/H\gtrsim 1$ and
a small $\mathrm{Ca}_f$ comparable to the experimental value with current computational resources. 
Nevertheless it is expected that as long as the thickness of the lubrication film does not reach the
regime where van der Waals forces are important, 
the effect of the lubrication film on the droplet dynamics can still be captured 
with simulation results at larger $\mathrm{Ca}_f$. 
At the end of this section an extensive discussion on the effect of surface tension will be given. 

\subsection{Effect of droplet horizontal radius on droplet dynamics}

\begin{figure}[tbp]
	\centering
	\includegraphics[width=0.96\columnwidth]{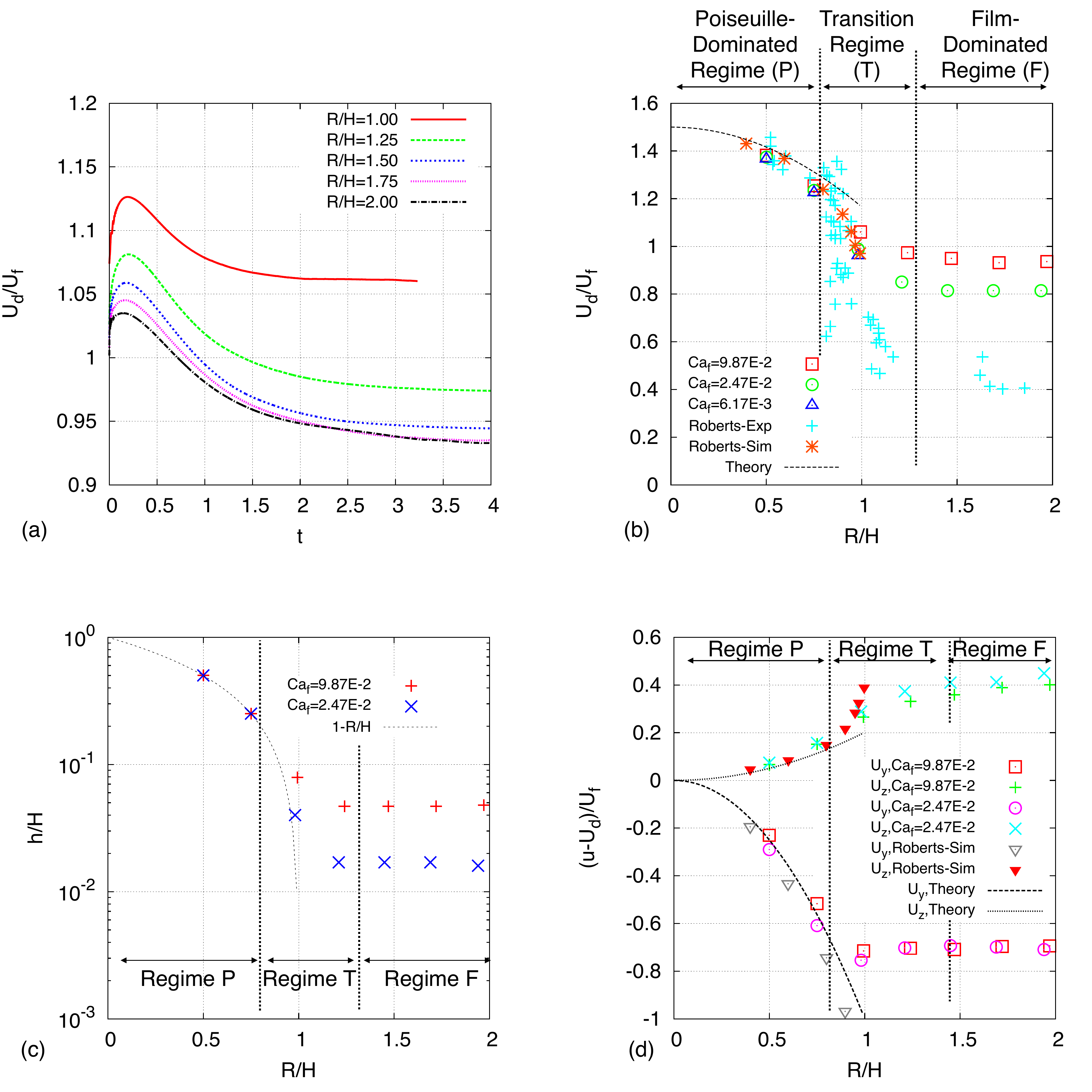} 
	\caption{(a) Temporal evolution of mean droplet velocity. (b) Terminal mean droplet velocity $U_d$, (c) thickness (height) of the gap (film) between the droplet interface and the top channel wall, denoted by $h$, and (d) relative $u$-velocities at the top and the side of the droplet ($U_y$ and $U_z$) as a function of the ratio between the droplet radius $R$ and half of the channel height $H$. The experimental and theoretical results are shown for comparison \cite{Roberts_2014a, Nadim_1991a,Hudson_2010a} .  }
	\label{fig:udvsR}
\end{figure}

The temporal evolution of the droplet velocity is shown in Fig.\ \ref{fig:udvsR}(a) 
for $\mathrm{Ca}_f=9.87\times 10^{-2}$ and $R/H$ varying from 1 to 2. 
When $R/H$ increases, $U_d$ decreases and in contrast, 
the time it takes for the droplet to reach its terminal velocity increases (while identical 
initial conditions are imposed). 
From $R/H=1$ to 1.25, $U_d$ decreases significantly (about 9\%). 
The decrease of $U_d$  with $R/H$ becomes more gradual for larger $R/H$. It is seen that 
the difference between $U_d$ for $R/H=1.75$ and $R/H=2$ becomes invisible as $t>2$. 

The simulation results of the droplet velocity as a function of $R/H$ are shown in Fig.\ \ref{fig:udvsR}(b). 
 The experimental and simulation results 
of Roberts \etal \cite{Roberts_2014a} and the theoretical results of Nadim and Stone \cite{Nadim_1991a}
for small $R/H$ are also shown for comparison. The present predictions of $U_d$ are 
in good agreement with both the theoretical and experimental results for $R/H<1$. 

For $R/H>1$ the simulation results of $U_d$ are larger than those measured in the experiment. 
The discrepancy is expected since smaller surface tension coefficients (or larger capillary numbers)
are considered in the simulation. 
Nevertheless the simulation results capture well the trend of variation of $U_d$ with $R/H$,
and the three different regimes (Poiseuille-dominated, transition, and film-dominated regimes) can 
be clearly seen. In particular, the variation of $U_d$  in the transition and the film-dominated regimes 
is much better illustrated. 
In the transition regime ($0.75\lesssim R/H \lesssim 1.25$), 
the droplet just overfills the channel forming a lubrication film. Correspondingly,
the droplet velocity decreases rapidly. 
In the film-dominated regime, for $R/H>1.5$, $U_d$ is observed to reach a plateau gradually. 
The plateau value of $U_d$ decreases with $\mathrm{Ca}_f$, 
approaching the experimental value which corresponds to a much smaller $\mathrm{Ca}_f$. 

The three different regimes can also be identified from the variation of film thickness with $R/H$ as 
shown in Fig.\ \ref{fig:udvsR}(c). In the Poiseuille-dominated regime, as the gap between the droplet and the top wall is relatively large, its thickness is not affected by the wall. The droplet remains close to a sphere, so $h/H$ closely follows $1-R/H$. As $R/H$ approaches 1, (in the transition regime,)  the film thickness starts to deviate from $1-R/H$ and becomes governed by the viscous stress and the surface tension. When $R/H\gg 1$, (in the film-dominated regime,) the film thickness becomes completely independent from the horizontal radius and becomes dependent on $\mathrm{Ca}_f$. 

At last the droplet circulation velocities in the three different regimes are shown in 
Fig.\ \ref{fig:udvsR}(d). The $u$-velocities at the top (the point with maximum $y$ on the droplet) 
and the side of the droplet (the point with minimum $z$ on the droplet)  are denoted as $U_y$
and $U_z$, respectively. The simulation results of Roberts \etal \cite{Roberts_2014a} 
and the theoretical results of Hudson \cite{Hudson_2010a} are shown for comparison. 
For $R/H \ll 1$, namely in the Poiseuille-dominated regime,  
$U_y$ and $U_z$ vary as parabolic functions with respect to $R/H$. 
The present predictions of $U_y$ are in good agreement with both the previous 
numerical results and the theoretical prediction. 
In the film-dominated regime ($R/H \gg 1$), it is observed that both $U_y$ and $U_z$ 
reach constant values, similar to $U_d$. 
It is also interesting to notice that the relative velocities $U_y-U_d$ and $U_z-U_d$ for 
different $\mathrm{Ca}_f$ seem to overlap. 

\subsection{A simple model for droplet dynamics}
An interesting observation from the simulation results is that
the velocity of a 3D droplet can vary in a much wider range than the velocity of a 2D droplet. 
In particular, the droplet velocity in the film-dominated regime is observed to be generally 
smaller than the average inflow velocity. This low droplet velocity has also been observed in experiments 
\cite{Roberts_2014a, Huerre_2015a}. 

From Fig.\ \ref{fig:udvsR}(b) it can be seen that 
when $R/H$ increases from 0.5 to 2, the droplet velocity $U_d$ 
for $\mathrm{Ca}_f=2.47\times 10^{-2}$ decreases from 1.4 to 0.8. 
In this section, we propose a simple model of mass fluxes balance to qualitatively illustrate 
the droplet mobility in the film-dominated regime and the resulting 3D flow patterns. 

\label{sec:model}
\begin{figure}[tbpp]
	\centering
	\includegraphics[width=0.85\columnwidth]{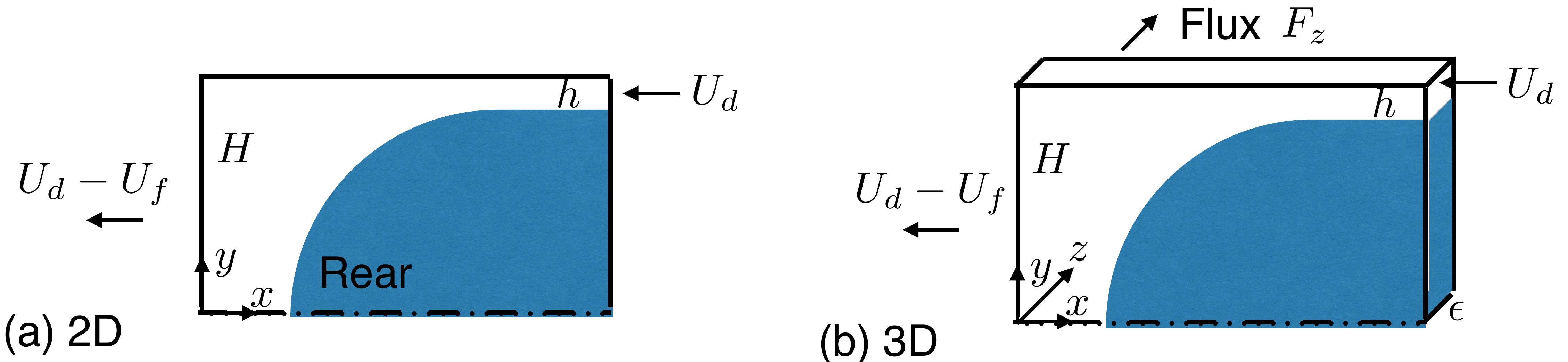} 
	\caption{Schematics of mass fluxes budget for a droplet in (a) 2D and (b) 3D channels. }
	\label{fig:mass_cons}
\end{figure}

In the 2D planar (or axisymmetric) geometry the ambient fluid which enters the channel 
can only move along the flow direction, and thus must push the droplet to move faster than 
the average inflow velocity to ensure mass conservation. 
If $h$ is given the corresponding value of $U_d$ can be obtained  by 
the balance of mass fluxes \cite{Bretherton_1961a}. 
Taking a control volume (of zero thickness) near the rear of the droplet as shown in Fig.\ \ref{fig:mass_cons}(a), 
the mass fluxes are identical to zero on both the $y$ and $z$ boundaries and thus the fluxes 
from the left and right must cancel. 
Considering a bubble with large $R/H$ and small $h/H$, 
with a stress-free interface (and thus a constant velocity within the film), 
the relation between $h$ and $U_d$ can be expressed as
\begin{equation}
	  \frac{U_d-U_f}{U_d}\frac{H}{h} = 1\, . 
	\label{eq:2D_mass_cons}
\end{equation}
It is seen that $U_d > U_f$ for non-zero $h$. For small $h/H$ the excess of droplet velocity 
is small and the droplet velocity $U_d$ is close to the average inflow velocity $U_f$. 

In the 3D analogue, since the ambient fluid can flow around the droplet 
from the sides (or edge) of the droplet, the mass fluxes balance and the flow pattern 
becomes more complex. 
Similarly to the 2D case we look at a thin layer of small thickness $\epsilon \ll h$ 
along the symmetric plane ($z\in (0,\epsilon)$) near the rear of the droplet. 
As there is zero flux on the symmetric plane $z=0$, 
the transverse mass flux is only non-zero at the plane $z=\epsilon$,  normalized by the flux from the right, 
\ie, $U_d h \epsilon$, which can be expressed as 
\begin{equation}
	 F_{z,rear} = 1 - \frac{U_d-U_f}{U_d}\frac{H}{h} \, , 
	\label{eq:3D_mass_cons}
\end{equation}
assuming the variation of $U_f$ in $z\in (0,\epsilon)$ is negligible. 
The transverse flux $F_{z,rear}$ is not uniformly distributed on the plane $z=\epsilon$, instead it is concentrated near the stagnation point at the rear of the droplet. 
A similar analysis can be performed for the front of the droplet and it can be shown 
that $F_{z,rear}$ and $F_{z,front}$ have the same magnitude but different signs, 
which creates a circulation around the edge of the droplet.

For a given $h$, $F_{z,rear}$ is related to $U_d$ through mass conservation, however
the fluxes balance is not sufficient to determine $U_d$. This extra spatial degree
of freedom of mass flux thus allows a much wider variation of droplet velocity.
In contrast to the 2D case, where $U_d$ is always slightly larger than $U_f$, in the 3D case $U_d/U_f$ has been 
observed to vary from as low as $0.1$ to about 2 in experiments \cite{Park_1994a, Huerre_2015a}. 

Furthermore the two scenarios of $U_d>U_f$ and $U_d<U_f$ correspond to 
two qualitatively different three-dimensional flow patterns over the droplet. 
Assuming $h\ll H$, it can be shown that $F_{z,rear}<0$ when $U_d>U_f$. 
The ``pure" bubble experiment by Park \etal \cite{Park_1994a} 
falls in this scenario of high-migration-velocity (HMV).
In the droplet reference frame the sweeping motion of the ambient fluid 
on the edge of the droplet is from the front to the rear. 
Combined with the fact that the flow on the top of the droplet is always from the front to the rear, 
this means that the overall flow in the droplet reference frame is in the rear-to-front direction. 

On the other hand the scenario of low-migration-velocity (LMV), \ie, $U_d<U_f$ and thus $F_{z,rear} > 0$, 
is very different.
The droplets in the present simulation and in the experiments of Roberts \etal \cite{Roberts_2014a}
and Huerre \etal \cite{Huerre_2015a} fall in this category. 
The positive transverse flux near the droplet rear ($F_{z,rear}>0$) indicates that 
the fluid actually goes around the droplet from the rear to the front in the droplet reference frame.  
As a consequence, the relative flows on the top and the side of the droplets are in opposite directions, 
(which are also illustrated in Fig.\ \ref{fig:udvsR}(d)). 
A complex interfacial flow sweeping over the droplet is then formed, 
in which the flow (in the droplet reference frame) first goes from the front to the rear 
on the top of the droplet then returns from the rear back to the front on the edge
of the droplet. (A visualization of this interfacial flow will be presented in Fig.\ \ref{fig:interfacial_vel-vector_dropframe}.)

\subsection{Effect of three-dimensional flow structure on droplet dynamics}

\subsubsection{Streamlines and flow circulation}
\begin{figure}[tbp]
	\centering
	\includegraphics[width=1\columnwidth]{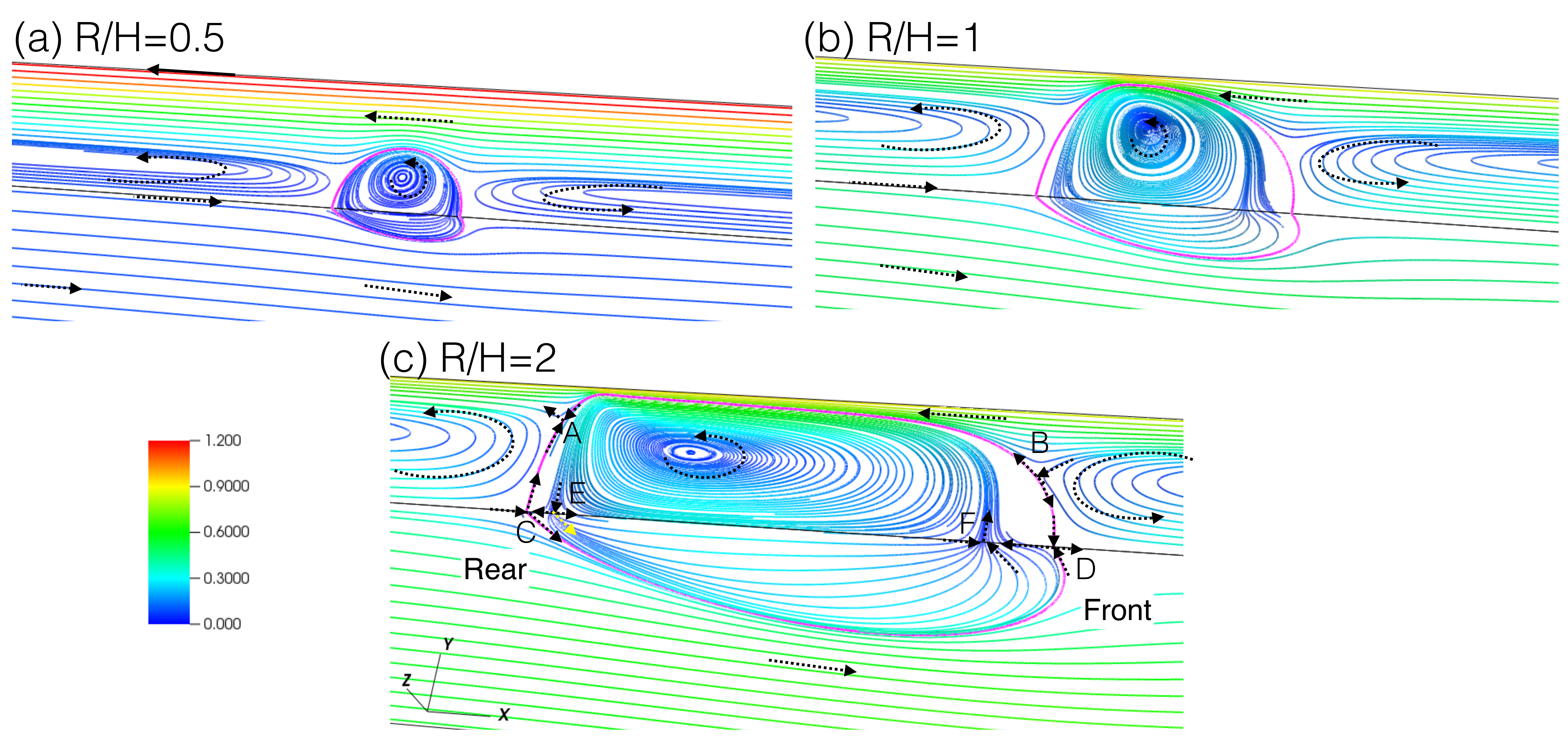} 
	\caption{Streamlines inside and outside of the droplet in the droplet reference frame for $\mathrm{Ca}_f=9.87\times 10^{-2}$  and $R/H=0.5$, 1 and 2. The color of the streamlines 
	denotes the corresponding relative velocity magnitude. The pink line indicates the droplet interface.}
	\label{fig:streamlines}
\end{figure}

\begin{figure}[tbp]
	\centering
	\includegraphics[width=1.0\columnwidth]{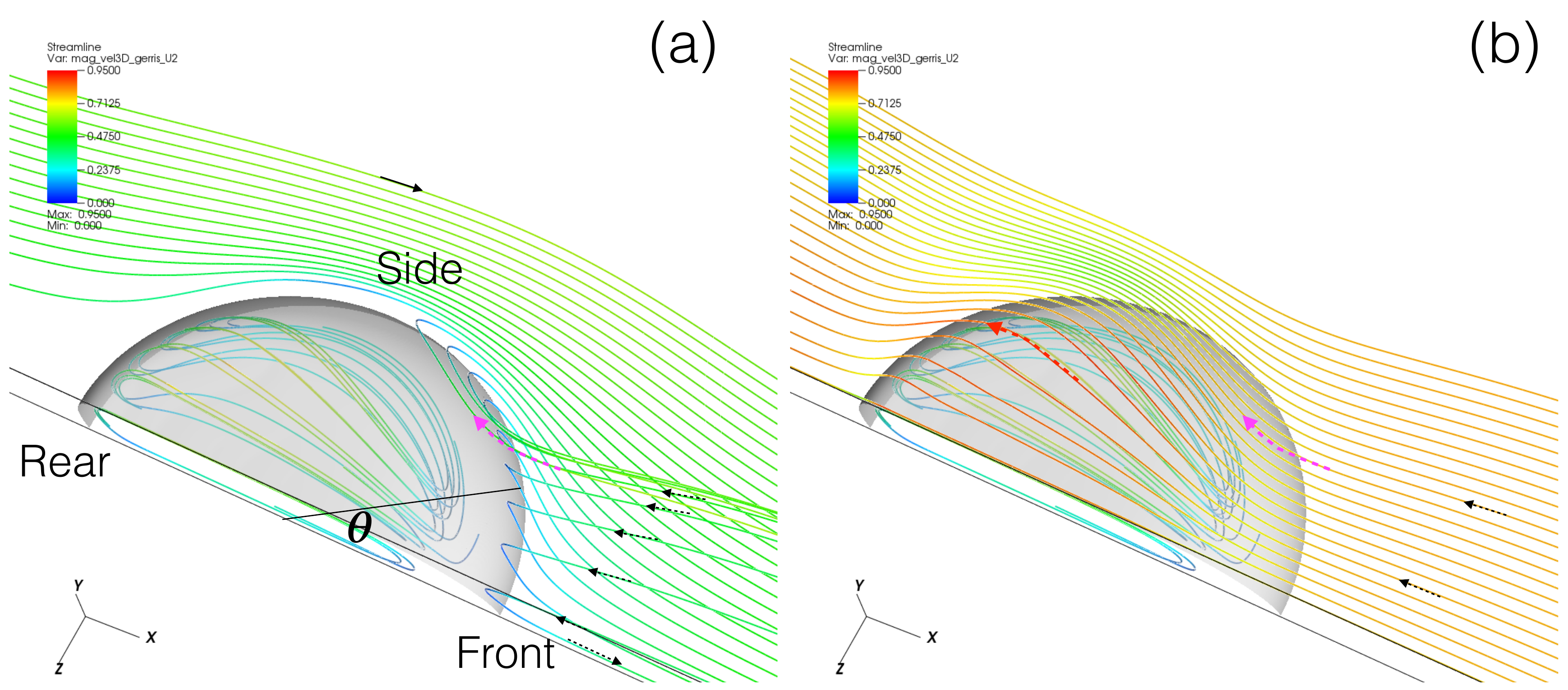} 
	\caption{Streamlines for $\mathrm{Ca}_f=9.87\times 10^{-2}$  and $R/H=2$
	in the droplet reference frame: 
	(a) Internal circulation and external circulation near the front of the droplet. (b) Flow 
	within the film between the droplet and the wall and the internal circulation inside the droplet.
	The color of the streamlines denotes the magnitude of the relative velocity. 
	The angular coordinate $\theta$ is defined in the droplet reference frame with the origin 
	located at the center of the horizontal cross-section.}
	\label{fig:streamlines_3D}
\end{figure}

The three-dimensional flow structure inside and outside of the droplet is shown 
in Fig.\ \ref{fig:streamlines} for $\mathrm{Ca}_f=9.87\times 10^{-2}$  and $R/H=0.5$, 1 and 2. 
The streamlines on the two symmetric planes $z=0$ and $y=0$ are plotted in the droplet reference frame 
and colored by the relative velocity magnitude. 
As the droplet moves faster for smaller droplets the relative velocity magnitude is larger for 
smaller $R/H$. The flow pattern for $R/H<1$ here is also consistent with the simulation results 
of Roberts \etal \cite{Roberts_2014a}.

It is interesting to compare the 3D flow pattern for the droplet with $R/H=2$ to its 2D analogue shown 
in Fig.\ \ref{fig:2d_results}. In the vertical plane $z=0$ the ambient fluid is moving from right to left at the top ($y=H$) and in a reverse direction 
at the center plane ($y=0$). As a result two external circulations develop outside of the droplet, 
similar to the 2D analogue shown in Fig.\ \ref{fig:2d_results}(a). 
The external flow will also drive internal circulation inside the droplet. 
Similarly to the 2D case, six stagnation points (A, B, C, D on the droplet interface 
and E, F inside the droplet) can be seen, however the stagnation flow around these points
becomes three dimensional. For example in the 2D case, the flow reaches 
the rear stagnation point C and can only turn toward point A
and then turn again to the left and leave the domain, forming a counter-clockwise external circulation, 
see Fig.\ \ref{fig:2d_results}(a). In 3D the fluid has an extra degree of freedom 
and can go in the $z$ direction around the droplet toward the stagnation point D, 
see the streamlines on the $x$-$z$ plane in Fig.\ \ref{fig:streamlines}(c). 
This transverse fluid motion near the stagnation point C is characterized by 
$ F_{z,rear} $ in the model presented in section \ref{sec:model}.
Due to the 3D flow structure the minor internal circulations (A-E-C-A and B-D-F-B) 
in the 2D case are invisible here. 

The transverse fluid motion significantly affects the external and internal circulating flow. 
As shown in Fig.\ \ref{fig:streamlines_3D}(a), 
the 3D streamlines inside and outside of the droplet become very complicated 
and are very dependent on the angular coordinate $\theta$.
(Here $\theta$ is defined as the angle between the normal 
to the interface in the $x$-$z$ plane and the $x$ axis.)
For $\theta=0$ the circulation is perfectly aligned with the symmetric $x$-$y$ plane, 
and the flow pattern is similar to the 2D case. 
As $\theta$ increases the external circulation is twisted in a complex manner. 
The incoming flows for nonzero $\theta$ near the top of the channel 
are observed to turn toward the center plane $z=0$, (indicated by black arrows in Fig.\ \ref{fig:streamlines_3D}(a)). 
As the flow approach the droplet interface, it turns to the side to adjust to the droplet shape, 
(see the magenta arrow). 
At about $\theta \approx 60$ degrees the streamlines outside of the droplet 
are almost tangential to the droplet interface. 
Therefore, in contrast with the typical quasi-2D flows in a Hele--Shaw cell, 
droplet migration in the microchannel generates a fully 3D flow. 

\subsubsection{Velocity field on the droplet interface}
\begin{figure}[tbp]
	\centering
	\includegraphics[width=0.99\columnwidth]{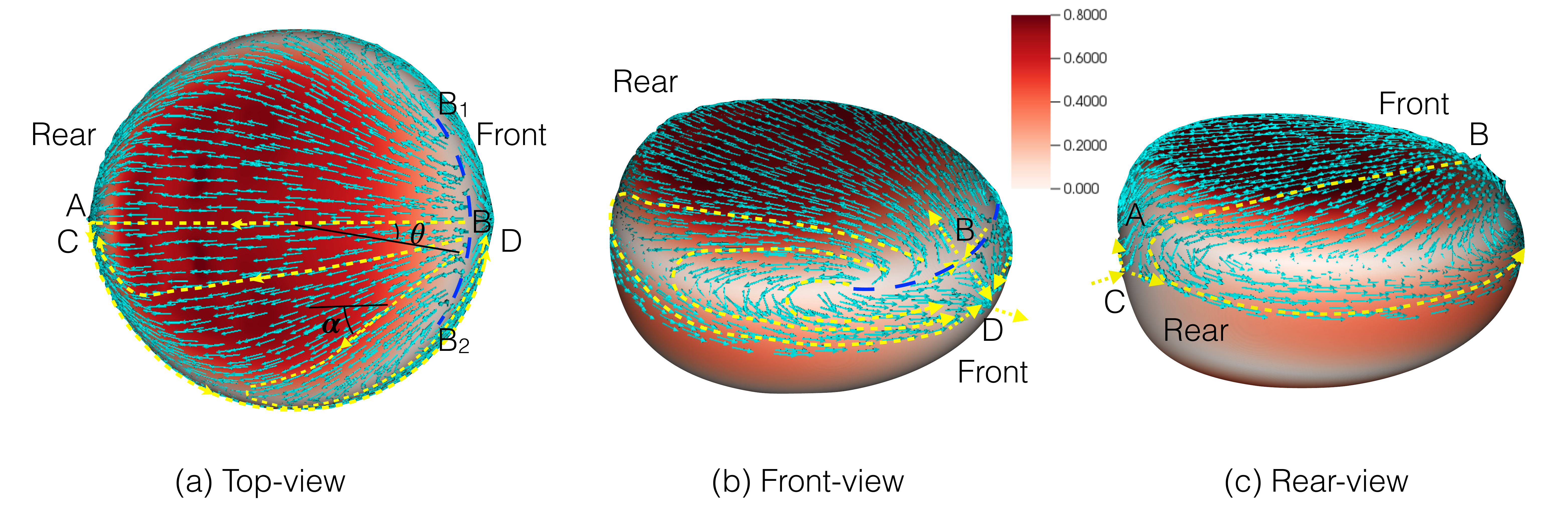} 
	\caption{(a) Top-  (b) front- and (c) rear-views of relative velocity in the droplet reference frame 
	$\{u-U_d,v,w\}$ on the droplet interface for $\mathrm{Ca}_f=9.87\times 10^{-2}$  
	and $R/H=0.5$, 1 and 2. The color denotes the magnitude of the relative velocity. }
	\label{fig:interfacial_vel-vector_dropframe}
\end{figure}

The vectors of the interfacial velocity in the droplet reference frame are shown 
in Fig.\ \ref{fig:interfacial_vel-vector_dropframe}. 
The color of the interface indicates the relative velocity magnitude. 
Regions with white color indicate zero velocity, \ie, stagnation regions on the droplet interface. 
The four stagnation points (A, B, C, D) on the droplet interface and 3D stagnation flows 
in the vicinity can be clearly seen. 

Driven by the flow within the film, the interfacial flow on the top of the droplet moves to the left; 
while the flow on the edge near the center plane $y=0$ moves in the opposite direction. 
An interesting and complex interfacial flow is then formed on the droplet interface, which is highlighted 
with yellow dashed lines in Fig.\ \ref{fig:interfacial_vel-vector_dropframe}. 
At first the flow starts from the vicinity of stagnation point B (indicated by the blue dashed line 
$\mathrm{B}_1$$\mathrm{B}_2$ ), and goes toward the rear.  
It is seen that the velocity vectors on the top interface are not parallel to the $x$ axis, instead they turn  toward the side of the droplet. 
The angle between the top interfacial flow and the $-x$ axis is denoted by $\alpha$ here, 
and it can be seen that $\alpha$ varies significantly with the angle $\theta$ 
of the relative flow near the wall (in the droplet reference frame). 
For $\theta=0$, $\alpha=0$, but then $\alpha$ increases significantly with $\theta$ up to about 45 degrees. 
Then top interfacial flow is observed to curve back toward the center plane ($z=0$) before reaching stagnation point A. 
After reaching point A the flow reverses its direction and moves toward the front stagnation point D 
along the edge of the droplet. For the top interfacial flow leaving from $\mathrm{B}_1$$\mathrm{B}_2$
with a large $\theta$, the flow may not get a chance to turn back to stagnation point A before returning
to the front. As a result a ring of zero interfacial velocity (the region in white color) is formed on the edge 
of the top where the interfacial flow reverses direction. 
Near the rear of the droplet the expected up-going interfacial flow from stagnation points C to A 
is not seen due to the strong interfacial flow along the edge. The stagnation flows near
C and A seem to merge, forming a stagnation area at the droplet rear. 

Furthermore, the streamlines in Fig.\ \ref{fig:streamlines_3D}(b) show that
the flow within the lubrication film is influenced by the complex interfacial flow on the droplet interface. 
Similar to the interfacial flow on the top of droplet (see Fig.\ \ref{fig:interfacial_vel-vector_dropframe}),
when the flow passes through the lubrication film, the streamlines off the center plane $z=0$ 
also first turn toward the side and then return back to the streamwise direction.

\subsection{Configuration and dynamics of the lubrication film}

\begin{figure}[tbp]
	\centering
	\includegraphics[width=0.85\columnwidth]{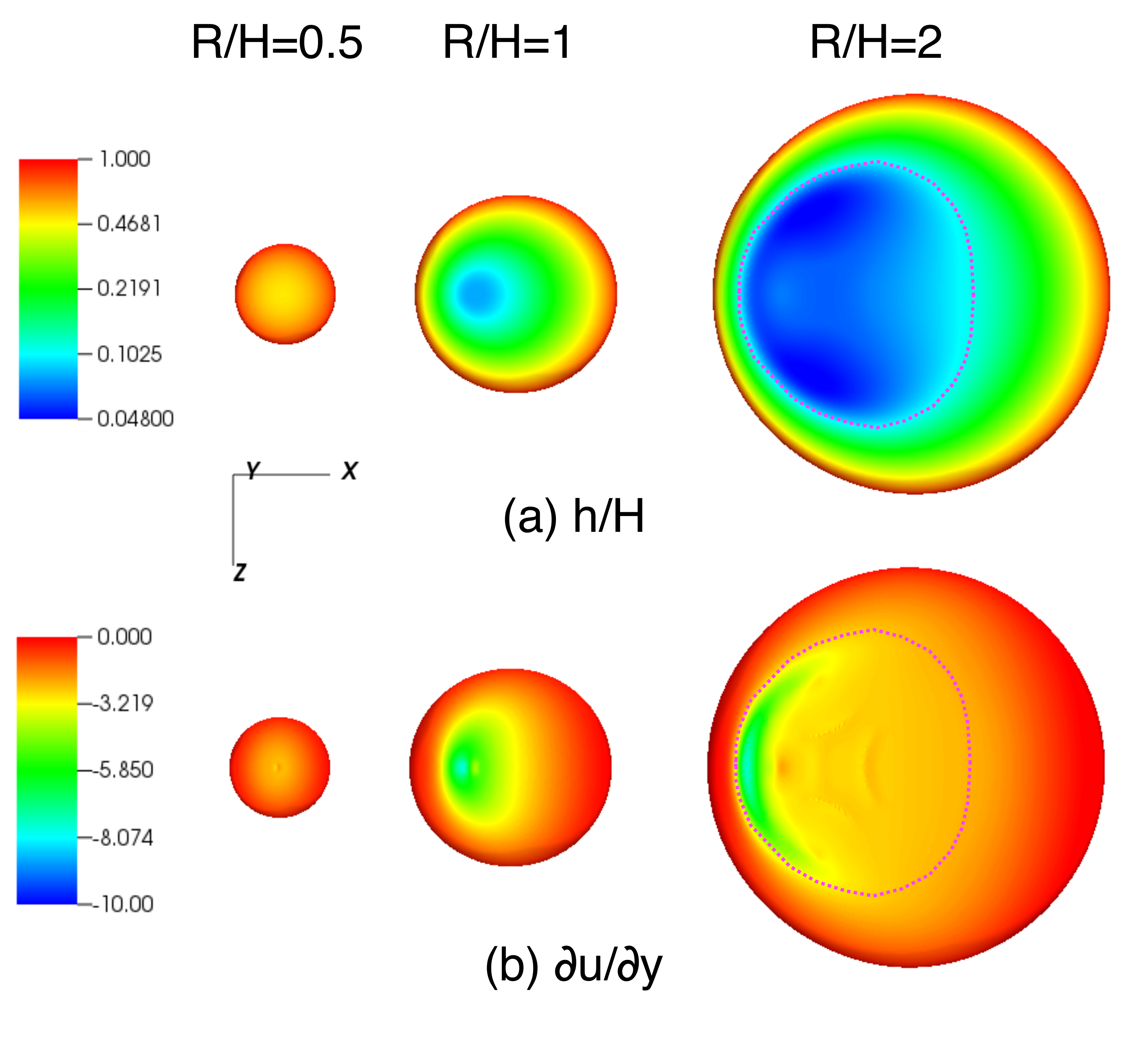} 
	\caption{(a) Thickness of the lubrication film between the interface and the wall $h/H$ and 
	(b) shear $\partial u/\partial y$ on the interface for $\mathrm{Ca}_f=9.87\times 10^{-2}$  
	and $R/H=0.5$, 1 and 2.  The film for $R/H=2$ is highlighted by pink dotted lines.  }
	\label{fig:interface_height_shear}
\end{figure}

As the film between the droplet and the wall serves to ``lubricate" the droplet, 
when the film thickness decreases (for example due to increasing surface tension) 
it is expected the droplet will slow down correspondingly.  
The increase of droplet velocity with the film thickness has been confirmed in the 
experiments by  Huerre \etal \cite{Huerre_2015a}. 
To better understand the effect of the lubrication film on the droplet dynamics, we plot 
the thickness of the lubrication film and the shear on the droplet top interface 
in Fig.\ \ref{fig:interface_height_shear}
for $\mathrm{Ca}_f=9.87\times 10^{-2}$  and $R/H=0.5$, 1 and 2. 

As $R/H$ is small the gap between the droplet and the wall is large 
and thus strictly speaking there exists no film, the term is retained only for convenience. 
When $R/H$ is larger or comparable to unity a thin lubrication film with small thickness can be clearly seen. 
In particular for $R/H=2$ the film area (highlighted by the pink dotted line) 
becomes quite large compared to the cross-section of the droplet at the $y=0$ plane. 
Instead of being centered, the lubrication film is shifted toward the rear
(see Fig.\ \ref{fig:interfacial_vel-vector_dropframe}).
The overall film exhibits a ``peach" shape.

Values of the shear $\partial u/\partial y$ on the droplet interface 
are shown in Fig.\ \ref{fig:interface_height_shear}(b). When $R/H$ is small the droplet is away from 
the top wall and the shear on the top of the droplet is indeed quite small. 
The shear increases dramatically as $R/H$ becomes comparable or larger than unity. 
For $R/H=2$ it is observed that the shear on the droplet is not uniformly distributed 
and the large negative shear is located near the rear. The region of high shear has
a croissant shape, and minimum shear rate reaches about $(\partial u/\partial y)_{\min}=-8$ 
(in dimensionless unit), the magnitude of which is more than twice larger than the minimum shear 
in the ambient 2D Poiseuille flow, \ie $(\partial u/\partial y)_{\min, \mathrm{Poiseuille}}=-3$. 
It is obvious that the strong shear stress induced by the film is closely related to the low 
mobility of droplets with larger $R/H$. 

\subsection{Effect of capillary number on film and droplet dynamics}
As mentioned at the beginning of this section, the capillary numbers considered
in the present 3D simulations are larger than that in experiments. 
To justify that the simulation results presented above at least qualitatively reflect the correct 
flow physics of droplet migration in a Hele--Shaw cell, 
the capillary number is varied to investigate its effect on the results. 
From the droplet velocity previously shown in Fig.\ \ref{fig:udvsR}(b), the change in $U_d$ 
is invisibly small for $R/H=0.5$ 
as $\mathrm{Ca}_f$ varies from $9.87\times 10^{-2}$ to $6.17\times 10^{-3}$.
Though smaller capillary numbers are used, the surface tension is still sufficient 
to keep the droplet spherical and the droplet velocity is then mainly determined by the viscous stress. 
The droplet velocity obtained in simulations for $\mathrm{Ca}_f=9.87\times 10^{-2}$, 
which is two orders of magnitude larger than that in experiments, still agrees quite well with 
the experimental and theoretical results. 
On the other hand, as $R/H\gtrsim 1$, the effect of the capillary number on the droplet velocity 
becomes much stronger. As shown in Fig.\ \ref{fig:udvsR}(b) and (c), the droplet velocity and film thickness 
decrease by about $12.2\%$ and $53.3\%$ for $R/H=2$, when $\mathrm{Ca}$ decreases from 
$9.87\times 10^{-2}$ to $2.47\times 10^{-2}$. 

\begin{figure}[tbp]
	\centering
	\includegraphics[width=0.85\columnwidth]{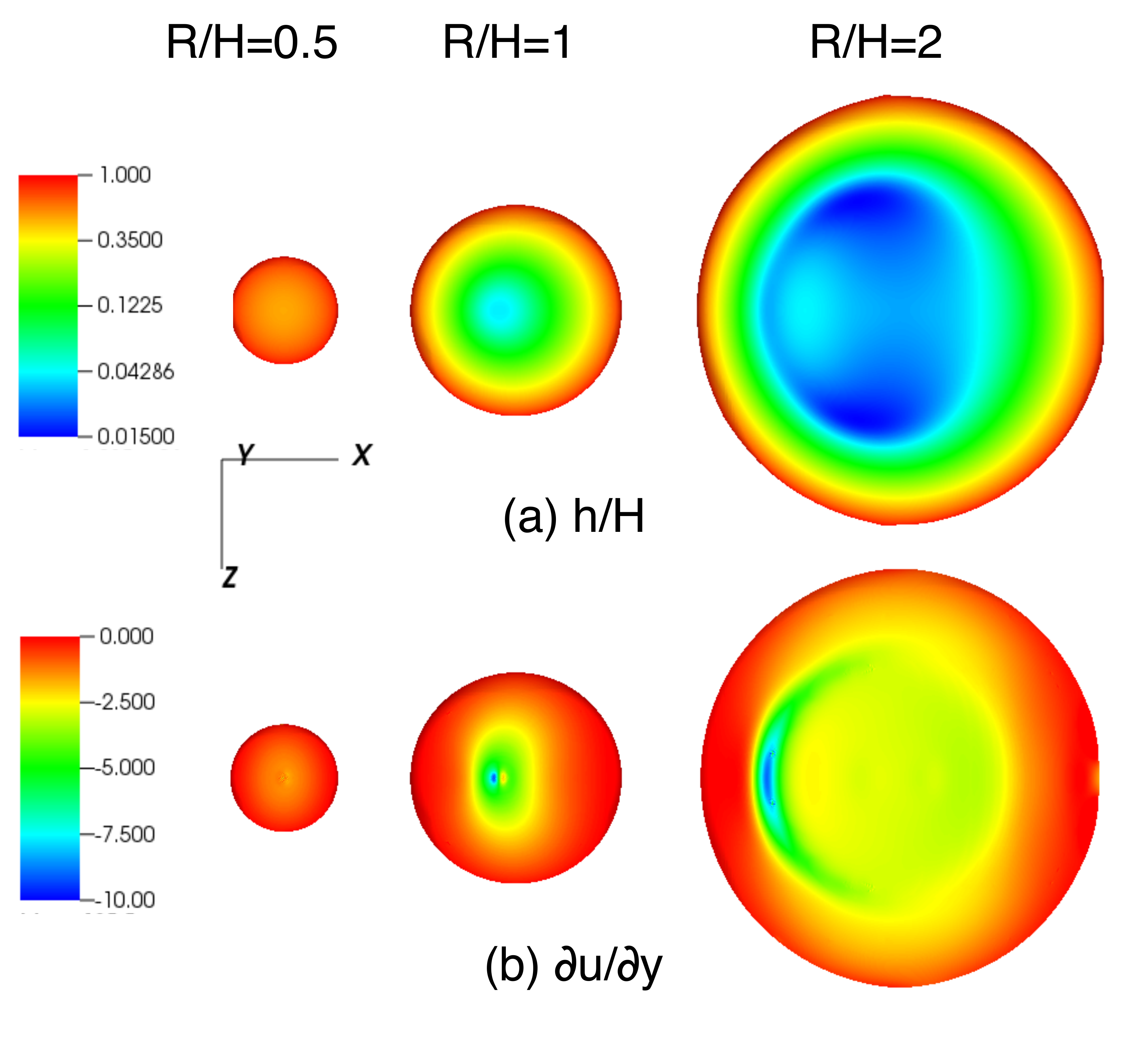} 
	\caption{(a) Thickness of the lubrication film between the interface and the wall $h/H$ and (b) shear $\partial u/\partial y$ on the interface for $\mathrm{Ca}_f=2.47\times 10^{-2}$  and $R/H=0.5$, 1 and 2.  }
	\label{fig:interface_height_shear_SmallST}
\end{figure}

The film thickness and shear on the droplet interface 
for $\mathrm{Ca}_f=2.47\times 10^{-2}$  and $R/H=0.5$, 1 and 2 are shown 
in Fig.\ \ref{fig:interface_height_shear_SmallST}. 
The surface tension here is four times that in Fig.\ \ref{fig:interface_height_shear} 
while the other parameters remain the same. 
(Correspondingly $\mathrm{Ca}_f$ in Fig.\ \ref{fig:interface_height_shear_SmallST}
is a quarter of that in Fig.\ \ref{fig:interface_height_shear}.)
When $R/H=0.5$ there is no difference in the gap thickness. 
For $R/H=2$ the film becomes significantly thinner as surface tension increases. (Therefore
different legends for film thickness are used in in Figs.\ \ref{fig:interface_height_shear}(a) and \ref{fig:interface_height_shear_SmallST}(a) for better visualization.)
The minimum film thickness for $\mathrm{Ca}_f=2.47\times 10^{-2}$ is about $h_{\min}=0.015H$ while 
for $\mathrm{Ca}_f=9.87\times 10^{-2}$, $h_{\min}=0.048H$. 
It is also observed that when $\mathrm{Ca}_f$ decreases, the lubrication film 
still shifts to the rear, however the shift becomes smaller and the droplet shape becomes 
more symmetric with regard to the $y$-$z$ plane. 
Furthermore, the overall ``peach" shape of the film remains, but the smallest film thickness
for $R/H=2$ seems to move to the two sides of the film. 

When $\mathrm{Ca}_f$ decreases the general distribution of the shear rate within the film is similar.
The high-shear region still holds a croissant shape. 
The minimum shear near the rear decreases 
to about $-10$ which is about 25\% larger than that for $\mathrm{Ca}_f=9.87\times 10^{-2}$ in magnitude. 
This indicates that the decreases of $\mathrm{Ca}_f$ results in a decrease of $h$ which in turn 
introduces a stronger shear drag on the droplet and reduces the droplet migration velocity. 

\begin{figure}[tbp]
	\centering
	\includegraphics[width=0.99\columnwidth]{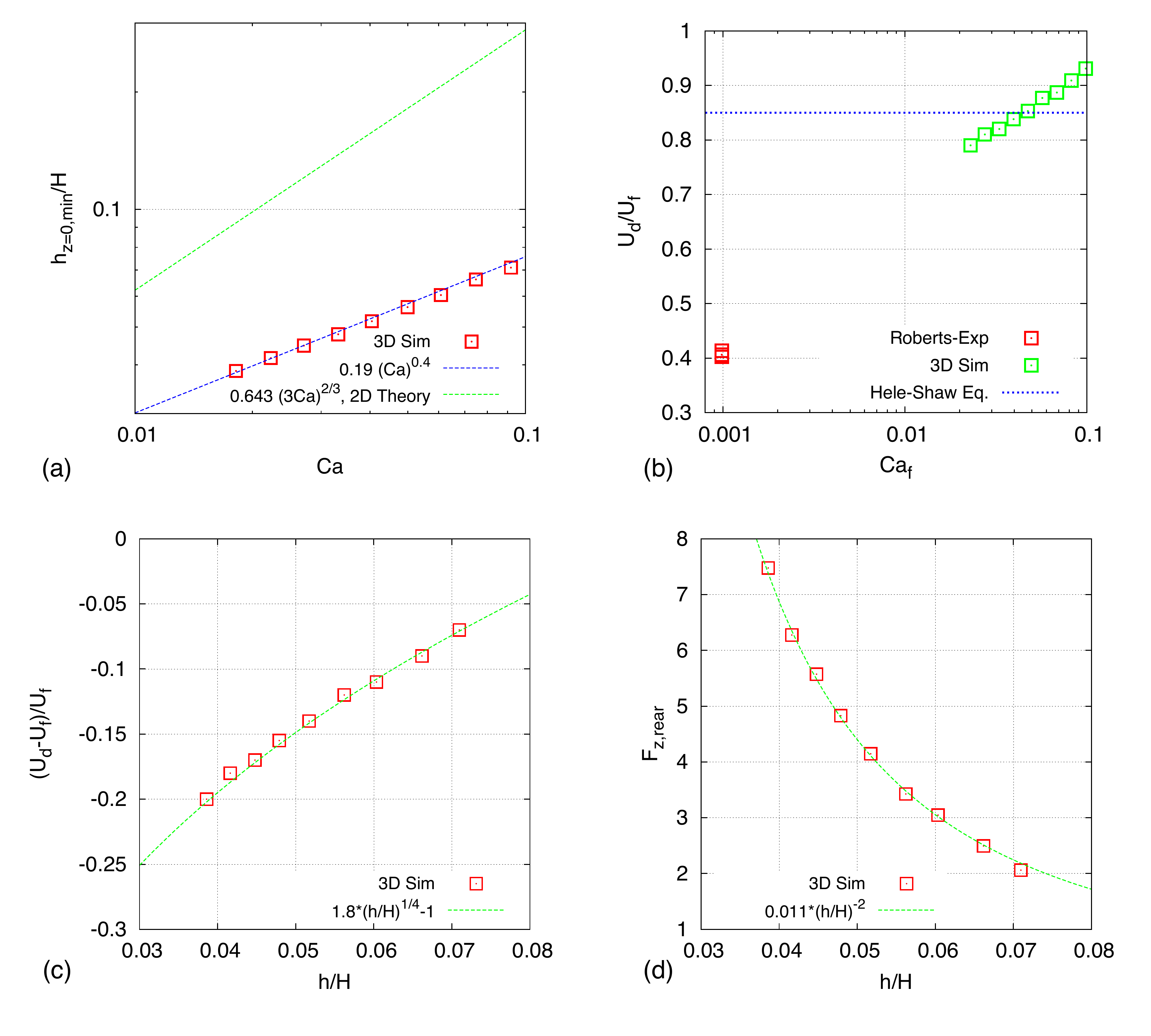} 
	\caption{(a) The minimum film thickness at the $z=0$ plane and (b) droplet velocity as functions of the capillary number. The droplet velocity obtained by the Hele--Shaw equations (Eq.\ \eqr{HS_drop_vel}) is also plotted for comparison. (c) Excess of droplet 
	velocity compared to the average inflow velocity and (d) Normalized transverse flux near the rear of the droplet 
	as a function of the film thickness. 
	}
	\label{fig:ud-h_vs_Ca}
\end{figure}

The minimal film thickness on the symmetric $x$-$y$ plane is plotted 
as a function of $\mathrm{Ca}$ in Fig.\ \ref{fig:ud-h_vs_Ca}(a), 
where $\mathrm{Ca}=\mu_f U_d/\sigma$ is the capillary number based on the droplet velocity $U_d$. 
The 2D theory, \ie, Eq.\ \eqr{scaling_2d_inviscid}, is also plotted for comparison. 
It is seen that $h_{z=0,\min}/H$ in the 3D case is much smaller than in the 2D case. 
For $\mathrm{Ca}=9.87\times 10^{-2}$, $h_{z=0,\min}/H$ in 3D is an order of magnitude smaller than in 2D, 
as already visualized in Figs.\ \ref{fig:2d_results} and \ref{fig:streamlines}. 
The difference between 3D and 2D film thickness seems to decrease with $\mathrm{Ca}$. 
The film thickness variation with $\mathrm{Ca}$ seems to follow a power law but the power coefficient 
is about 0.4 instead of 2/3 as in 2D. 

The droplet velocity $U_d$ decreases with $\mathrm{Ca}_f$ as shown in Fig.\ \ref{fig:ud-h_vs_Ca}(b). 
Here the capillary number based on the average inflow velocity is used following previous works 
\cite{Kopf-Sill_1988a,Park_1994a}. 
The droplet velocity obtained by the Hele--Shaw equations (Eq.\ \eqr{HS_drop_vel}), which is about 0.85, 
is also plotted for comparison.
The depth-averaged approximation misses the effect of the lubrication film on the droplet dynamics, 
and thus Eq.\ \eqr{HS_drop_vel} fails to capture the variation of the droplet velocity with capillary number. 
It is seen that the Hele--Shaw solution under-predicts $U_d/U_f$ for larger $\mathrm{Ca}_f$ 
and over-predicts $U_d/U_f$ for small $\mathrm{Ca}_f$. The droplet velocity only matches 
the Hele--Shaw solution at $\mathrm{Ca}_f\approx 0.05$. Similar observations have also been made by 
Zhu and Gallaire \cite{Zhu_2016a}. 
The range of $\mathrm{Ca}_f$ considered in the present simulation is an order of magnitude larger 
than the experimental value. It is hard to tell if the simulation results would match the experimental 
measurements \cite{Roberts_2014a} if the same $\mathrm{Ca}_f$ is used. Nevertheless
it seems like the computed $U_d$ would approach the experimental value as $\mathrm{Ca}_f$ decreases. 

As discussed previously the mass fluxes balance in 3D is substantially different from 2D. 
In the 2D case it is known that $(U_d-U_f)/U_f=h/H$ (for $h/H\ll1$). In 3D, 
$(U_d-U_f)/U_f$ becomes negative. It is shown that $U_d$ also decreases with $h/H$, 
but the decrease is non-linear and follows a power law about $(h/H)^{1/4}$ 
as indicated in Fig.\ \ref{fig:ud-h_vs_Ca}(c).

The simple model of transverse flux near the rear of the droplet (Eq.\ \eqr{3D_mass_cons})
can serve as a measure of the three-dimensional ``strength'' of the flow. The variation of $F_{z,rear}$
is plotted as a function of the film thickness in Fig.\ \ref{fig:ud-h_vs_Ca}(d). 
When $h$ decreases it is observed that $F_{z,rear}$ increases rapidly. 
As the film becomes thinner, the drag induced by the lubrication film becomes stronger and it is more 
difficult for the ambient fluid to push the droplet forward. As a result
the ambient fluid  entering the channel tends to go around the droplet from the transverse edge 
instead of pushing the droplet. Therefore it is quite clear that both the 3D flow effect and the lubrication film 
dynamics control  the mobility of a droplet in a Hele--Shaw cell. 

\subsection{Discussions of computational costs}
\begin{figure}[tbp]
	\centering
	\includegraphics[width=1.0\columnwidth]{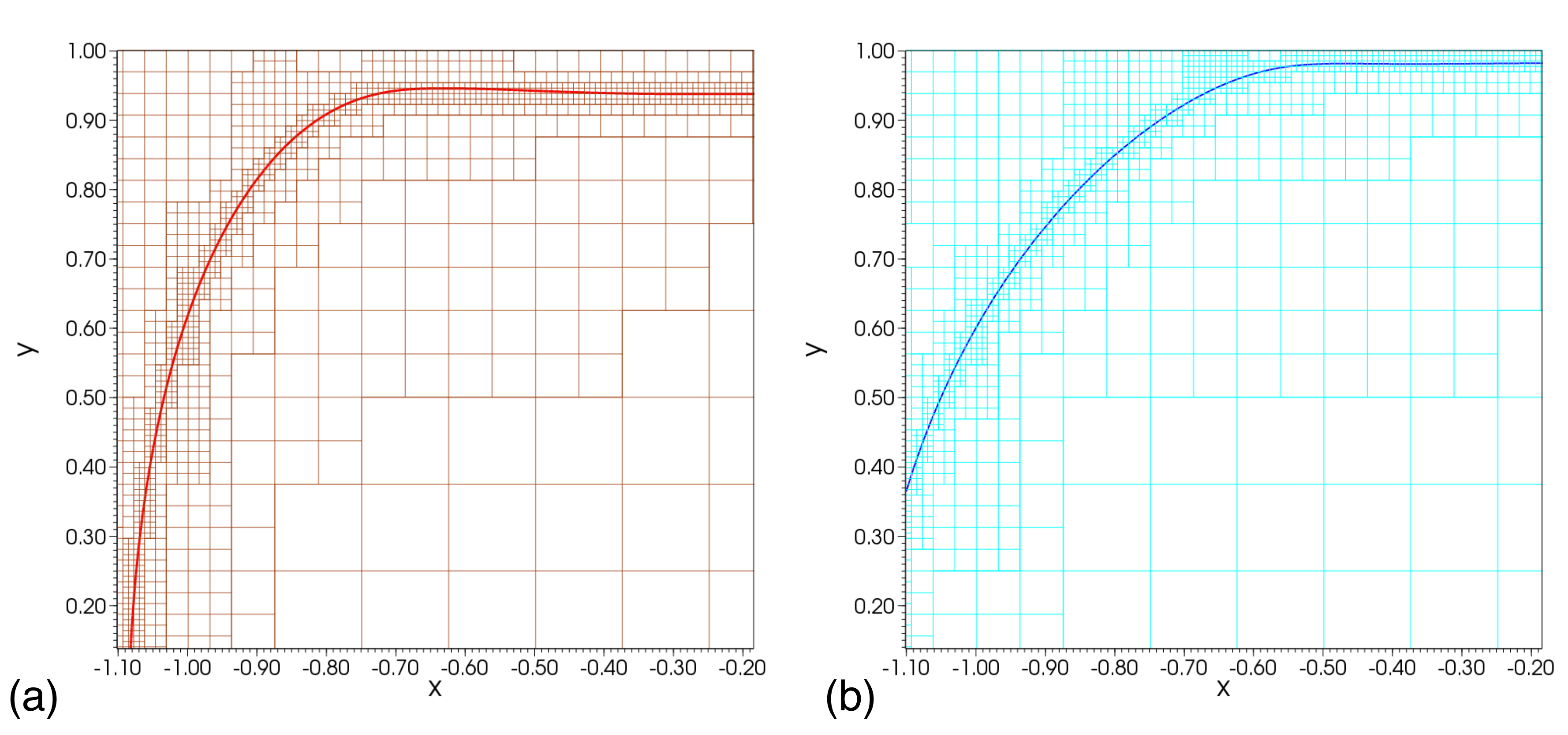} 
	\caption{The adaptive mesh on the plane $z=0$ for $R/H=2$  and (a) $\mathrm{Ca}_f=9.87\times 10^{-2}$
	and (b) $\mathrm{Ca}_f=2.47\times 10^{-2}$.  }
	\label{fig:2D-interface_with-mesh}
\end{figure}

The present simulations are performed on 8 to 24 processors for 3 to 25 days depending on cases. 
The adaptive meshes used for $\mathrm{Ca}_f=9.87\times 10^{-2}$  and $2.47\times 10^{-2}$ and $R/H=2$ 
at the plane $z=0$ are shown in Fig.\ \ref{fig:2D-interface_with-mesh}. The main constraint on the computational 
efficiency comes from the small time step which is in turn restricted by the surface tension (or capillary number), 
see Eq.\ \eqr{dt_surf_tension}. 
When surface tension increases four times, the time step is divided by two and namely 
the computational time doubles assuming the mesh is unchanged. 
However, as indicated in Fig.\ \ref{fig:2D-interface_with-mesh}(b) that if $\mathrm{Ca}_f$ 
is further decreased, the minimum mesh size in the adaptive mesh, \ie, $\Delta_{min}=H/128$, 
will not be sufficient and thus must be refined. Therefore, even with an adaptive mesh, it is too expensive to 
simulate droplets at capillary number comparable to experiments for large $R/H$. 
To perform 3D direct simulation of droplets at low  capillary number, 
novel numerical schemes to compute the surface tension are required. 
If the surface tension  term in Eq.\ \eqr{mom} can be treated in an implicit manner similar to what is 
done for the viscous term in Stokes flow (at low Reynolds number), then the time-step restriction due to the
explicit integration of the surface tension term can be lifted
(the authors are not aware of this type of scheme in the literature).
Nevertheless the development of an implicit treatment of surface tension is out of the scope of
the present paper and will be relegated to future work. 

\section{CONCLUSIONS}

The migration of a droplet in a Hele--Shaw cell has been investigated by 3D direct numerical simulations. We focused on the regime at low capillary number where the droplet remains circular in the horizontal plane. 
Parametric studies were performed by varying the droplet horizontal radius and the capillary number. 
For droplets with an horizontal radius $R$ larger than the half-height of the channel $H$, 
the droplet overfills the channel and a thin lubrication film is formed between the droplet and the wall. 
An adaptive two-phase flow solver is utilized for the simulations 
and mesh is locally refined to accurately resolve the thin film. 
The ratio $R/H$ has a significant impact on the droplet velocity and controls three regimes: 
the Poiseuille-dominated regime, the film-dominated regime, and the transition regime. 
The droplet velocity decreases dramatically with $R/H$ in the transition regime. 
In the film-dominated regime, the droplet horizontal radius is much larger than the channel height and 
the droplet holds a pancake-like shape.
Then the migration velocity is found to become independent of the radius and 
is mainly dictated by the capillary number. 
The simulation also shows that the droplet velocity in the film-dominated regime decreases with 
the capillary number and is generally lower than the average inflow velocity. 
The lubrication film dynamics and the three-dimensional flow structure seem to both
contribute to the low mobility of the droplet. In this low-migration-velocity scenario
the interfacial flow on the top of the droplet moves from front toward the rear in the droplet reference frame, 
but reverses its direction moving toward the front from the two sides of the the droplet. 
As the depth-averaged Hele--Shaw equations ignore the effect of the lubrication film on the droplet dynamics, 
their solution fails to capture the dependence of droplet velocity on capillary number. 
As the shear induced by the lubrication film has a strong influence on the droplet dynamics, 
the viscosity ratio can be an important 
parameter for the present problem, and we intend to investigate its influence in future work.

\section*{Acknowledgements}
This project has been supported by the the ANR TRAM project
(ANR-13-BS09-0011).  The simulations of this paper are conducted on our
laboratory cluster and on the CINES Occigen machine for which we gratefully 
acknowledge grant x20152b7325 from GENCI. 


\begin{thebibliography}{32}%
\makeatletter
\providecommand \@ifxundefined [1]{%
 \@ifx{#1\undefined}
}%
\providecommand \@ifnum [1]{%
 \ifnum #1\expandafter \@firstoftwo
 \else \expandafter \@secondoftwo
 \fi
}%
\providecommand \@ifx [1]{%
 \ifx #1\expandafter \@firstoftwo
 \else \expandafter \@secondoftwo
 \fi
}%
\providecommand \natexlab [1]{#1}%
\providecommand \enquote  [1]{``#1''}%
\providecommand \bibnamefont  [1]{#1}%
\providecommand \bibfnamefont [1]{#1}%
\providecommand \citenamefont [1]{#1}%
\providecommand \href@noop [0]{\@secondoftwo}%
\providecommand \href [0]{\begingroup \@sanitize@url \@href}%
\providecommand \@href[1]{\@@startlink{#1}\@@href}%
\providecommand \@@href[1]{\endgroup#1\@@endlink}%
\providecommand \@sanitize@url [0]{\catcode `\\12\catcode `\$12\catcode
  `\&12\catcode `\#12\catcode `\^12\catcode `\_12\catcode `\%12\relax}%
\providecommand \@@startlink[1]{}%
\providecommand \@@endlink[0]{}%
\providecommand \url  [0]{\begingroup\@sanitize@url \@url }%
\providecommand \@url [1]{\endgroup\@href {#1}{\urlprefix }}%
\providecommand \urlprefix  [0]{URL }%
\providecommand \Eprint [0]{\href }%
\providecommand \doibase [0]{http://dx.doi.org/}%
\providecommand \selectlanguage [0]{\@gobble}%
\providecommand \bibinfo  [0]{\@secondoftwo}%
\providecommand \bibfield  [0]{\@secondoftwo}%
\providecommand \translation [1]{[#1]}%
\providecommand \BibitemOpen [0]{}%
\providecommand \bibitemStop [0]{}%
\providecommand \bibitemNoStop [0]{.\EOS\space}%
\providecommand \EOS [0]{\spacefactor3000\relax}%
\providecommand \BibitemShut  [1]{\csname bibitem#1\endcsname}%
\let\auto@bib@innerbib\@empty
\bibitem [{\citenamefont {Stone}, \citenamefont {Stroock},\ and\ \citenamefont
  {Ajdari}(2004)}]{Stone_2004a}%
  \BibitemOpen
  \bibfield  {author} {\bibinfo {author} {\bibfnamefont {H.~A.}\ \bibnamefont
  {Stone}}, \bibinfo {author} {\bibfnamefont {A.~D.}\ \bibnamefont {Stroock}},
  \ and\ \bibinfo {author} {\bibfnamefont {A.}~\bibnamefont {Ajdari}},\
  }\bibfield  {title} {\enquote {\bibinfo {title} {Engineering flows in small
  devices: microfluidics toward a lab-on-a-chip},}\ }\href@noop {} {\bibfield
  {journal} {\bibinfo  {journal} {Annu.~Rev.~Fluid Mech.}\ }\textbf {\bibinfo
  {volume} {36}},\ \bibinfo {pages} {381--411} (\bibinfo {year}
  {2004})}\BibitemShut {NoStop}%
\bibitem [{\citenamefont {Fair}(2007)}]{Fair_2007a}%
  \BibitemOpen
  \bibfield  {author} {\bibinfo {author} {\bibfnamefont {R.~B.}\ \bibnamefont
  {Fair}},\ }\bibfield  {title} {\enquote {\bibinfo {title} {Digital
  microfluidics: is a true lab-on-a-chip possible?}}\ }\href@noop {} {\bibfield
   {journal} {\bibinfo  {journal} {Microfluid Nanofluid}\ }\textbf {\bibinfo
  {volume} {3}},\ \bibinfo {pages} {245--281} (\bibinfo {year}
  {2007})}\BibitemShut {NoStop}%
\bibitem [{\citenamefont {Teh}\ \emph {et~al.}(2008)\citenamefont {Teh},
  \citenamefont {Lin}, \citenamefont {Hung},\ and\ \citenamefont
  {Lee}}]{Teh_2008a}%
  \BibitemOpen
  \bibfield  {author} {\bibinfo {author} {\bibfnamefont {S.-Y.}\ \bibnamefont
  {Teh}}, \bibinfo {author} {\bibfnamefont {R.}~\bibnamefont {Lin}}, \bibinfo
  {author} {\bibfnamefont {L.-H.}\ \bibnamefont {Hung}}, \ and\ \bibinfo
  {author} {\bibfnamefont {A.~P.}\ \bibnamefont {Lee}},\ }\bibfield  {title}
  {\enquote {\bibinfo {title} {Droplet microfluidics},}\ }\href@noop {}
  {\bibfield  {journal} {\bibinfo  {journal} {Lab Chip}\ }\textbf {\bibinfo
  {volume} {8}},\ \bibinfo {pages} {198--220} (\bibinfo {year}
  {2008})}\BibitemShut {NoStop}%
\bibitem [{\citenamefont {Roberts}\ \emph {et~al.}(2014)\citenamefont
  {Roberts}, \citenamefont {Roberts}, \citenamefont {Nemer},\ and\
  \citenamefont {Rao}}]{Roberts_2014a}%
  \BibitemOpen
  \bibfield  {author} {\bibinfo {author} {\bibfnamefont {C.~C.}\ \bibnamefont
  {Roberts}}, \bibinfo {author} {\bibfnamefont {S.~A.}\ \bibnamefont
  {Roberts}}, \bibinfo {author} {\bibfnamefont {M.~B.}\ \bibnamefont {Nemer}},
  \ and\ \bibinfo {author} {\bibfnamefont {R.~R.}\ \bibnamefont {Rao}},\
  }\bibfield  {title} {\enquote {\bibinfo {title} {Circulation within confined
  droplets in {Hele--Shaw} channels},}\ }\href@noop {} {\bibfield  {journal}
  {\bibinfo  {journal} {Phys.~Fluids}\ }\textbf {\bibinfo {volume} {26}},\
  \bibinfo {pages} {032105} (\bibinfo {year} {2014})}\BibitemShut {NoStop}%
\bibitem [{\citenamefont {Nadim}\ and\ \citenamefont
  {Stone}(1991)}]{Nadim_1991a}%
  \BibitemOpen
  \bibfield  {author} {\bibinfo {author} {\bibfnamefont {A.}~\bibnamefont
  {Nadim}}\ and\ \bibinfo {author} {\bibfnamefont {H.~A.}\ \bibnamefont
  {Stone}},\ }\bibfield  {title} {\enquote {\bibinfo {title} {The motion of
  small particles and droplets in quadratic flows},}\ }\href@noop {} {\bibfield
   {journal} {\bibinfo  {journal} {Stud.~Appl.~Math.}\ }\textbf {\bibinfo
  {volume} {85}},\ \bibinfo {pages} {53--73} (\bibinfo {year}
  {1991})}\BibitemShut {NoStop}%
\bibitem [{\citenamefont {Hudson}(2010)}]{Hudson_2010a}%
  \BibitemOpen
  \bibfield  {author} {\bibinfo {author} {\bibfnamefont {S.~D.}\ \bibnamefont
  {Hudson}},\ }\bibfield  {title} {\enquote {\bibinfo {title} {Poiseuille flow
  and drop circulation in microchannels},}\ }\href@noop {} {\bibfield
  {journal} {\bibinfo  {journal} {Rheol.~Acta.}\ }\textbf {\bibinfo {volume}
  {49}},\ \bibinfo {pages} {237--243} (\bibinfo {year} {2010})}\BibitemShut
  {NoStop}%
\bibitem [{\citenamefont {Bretherton}(1961)}]{Bretherton_1961a}%
  \BibitemOpen
  \bibfield  {author} {\bibinfo {author} {\bibfnamefont {F.~P.}\ \bibnamefont
  {Bretherton}},\ }\bibfield  {title} {\enquote {\bibinfo {title} {The motion
  of long bubbles in tubes},}\ }\href@noop {} {\bibfield  {journal} {\bibinfo
  {journal} {J. Fluid Mech.}\ }\textbf {\bibinfo {volume} {10}},\ \bibinfo
  {pages} {166--188} (\bibinfo {year} {1961})}\BibitemShut {NoStop}%
\bibitem [{\citenamefont {Hodges}, \citenamefont {Jensen},\ and\ \citenamefont
  {Rallison}(2004)}]{Hodges_2004a}%
  \BibitemOpen
  \bibfield  {author} {\bibinfo {author} {\bibfnamefont {S.~R.}\ \bibnamefont
  {Hodges}}, \bibinfo {author} {\bibfnamefont {O.~E.}\ \bibnamefont {Jensen}},
  \ and\ \bibinfo {author} {\bibfnamefont {J.~M.}\ \bibnamefont {Rallison}},\
  }\bibfield  {title} {\enquote {\bibinfo {title} {The motion of a viscous drop
  through a cylindrical tube},}\ }\href@noop {} {\bibfield  {journal} {\bibinfo
   {journal} {J. Fluid Mech.}\ }\textbf {\bibinfo {volume} {501}},\ \bibinfo
  {pages} {279--301} (\bibinfo {year} {2004})}\BibitemShut {NoStop}%
\bibitem [{\citenamefont {Taylor}\ and\ \citenamefont
  {Saffman}(1959)}]{Taylor_1959a}%
  \BibitemOpen
  \bibfield  {author} {\bibinfo {author} {\bibfnamefont {G.~I.}\ \bibnamefont
  {Taylor}}\ and\ \bibinfo {author} {\bibfnamefont {P.~F.}\ \bibnamefont
  {Saffman}},\ }\bibfield  {title} {\enquote {\bibinfo {title} {A note on the
  motion of bubbles in a {Hele--Shaw} cell and porous medium},}\ }\href@noop {}
  {\bibfield  {journal} {\bibinfo  {journal} {Quart.~J.~Mech.~Appl.~Math.}\
  }\textbf {\bibinfo {volume} {12}},\ \bibinfo {pages} {265--279} (\bibinfo
  {year} {1959})}\BibitemShut {NoStop}%
\bibitem [{\citenamefont {Kopf-Sill}\ and\ \citenamefont
  {Homsy}(1988)}]{Kopf-Sill_1988a}%
  \BibitemOpen
  \bibfield  {author} {\bibinfo {author} {\bibfnamefont {A.~R.}\ \bibnamefont
  {Kopf-Sill}}\ and\ \bibinfo {author} {\bibfnamefont {G.~M.}\ \bibnamefont
  {Homsy}},\ }\bibfield  {title} {\enquote {\bibinfo {title} {Bubble motion in
  a {Hele--Shaw} cell},}\ }\href@noop {} {\bibfield  {journal} {\bibinfo
  {journal} {Phys.~Fluids}\ }\textbf {\bibinfo {volume} {31}},\ \bibinfo
  {pages} {18--26} (\bibinfo {year} {1988})}\BibitemShut {NoStop}%
\bibitem [{\citenamefont {Tanveer}(1986)}]{Tanveer_1986a}%
  \BibitemOpen
  \bibfield  {author} {\bibinfo {author} {\bibfnamefont {S.}~\bibnamefont
  {Tanveer}},\ }\bibfield  {title} {\enquote {\bibinfo {title} {The effect of
  surface tension on the shape of a {Hele--Shaw} cell bubble},}\ }\href@noop {}
  {\bibfield  {journal} {\bibinfo  {journal} {Phys.~Fluids}\ }\textbf {\bibinfo
  {volume} {29}},\ \bibinfo {pages} {3537--3548} (\bibinfo {year}
  {1986})}\BibitemShut {NoStop}%
\bibitem [{\citenamefont {Tanveer}(1987)}]{Tanveer_1987a}%
  \BibitemOpen
  \bibfield  {author} {\bibinfo {author} {\bibfnamefont {S.}~\bibnamefont
  {Tanveer}},\ }\bibfield  {title} {\enquote {\bibinfo {title} {New solutions
  for steady bubbles in a {Hele--Shaw} cell},}\ }\href@noop {} {\bibfield
  {journal} {\bibinfo  {journal} {Phys.~Fluids}\ }\textbf {\bibinfo {volume}
  {30}},\ \bibinfo {pages} {651--658} (\bibinfo {year} {1987})}\BibitemShut
  {NoStop}%
\bibitem [{\citenamefont {Saffman}\ and\ \citenamefont
  {Tanveer}(1989)}]{Saffman_1989a}%
  \BibitemOpen
  \bibfield  {author} {\bibinfo {author} {\bibfnamefont {P.~F.}\ \bibnamefont
  {Saffman}}\ and\ \bibinfo {author} {\bibfnamefont {S.}~\bibnamefont
  {Tanveer}},\ }\bibfield  {title} {\enquote {\bibinfo {title} {Prediction of
  bubble velocity in a {Hele--Shaw} cell: Thin film and contact angle
  effects},}\ }\href@noop {} {\bibfield  {journal} {\bibinfo  {journal}
  {Phys.~Fluids}\ }\textbf {\bibinfo {volume} {1}},\ \bibinfo {pages}
  {219--223} (\bibinfo {year} {1989})}\BibitemShut {NoStop}%
\bibitem [{\citenamefont {Park}, \citenamefont {Maruvada},\ and\ \citenamefont
  {Yoon}(1994)}]{Park_1994a}%
  \BibitemOpen
  \bibfield  {author} {\bibinfo {author} {\bibfnamefont {C.-W.}\ \bibnamefont
  {Park}}, \bibinfo {author} {\bibfnamefont {S.~R.~K.}\ \bibnamefont
  {Maruvada}}, \ and\ \bibinfo {author} {\bibfnamefont {D.-Y.}\ \bibnamefont
  {Yoon}},\ }\bibfield  {title} {\enquote {\bibinfo {title} {The influence of
  surfactant on the bubble motion in {Hele--Shaw} cells},}\ }\href@noop {}
  {\bibfield  {journal} {\bibinfo  {journal} {Phys.~Fluids}\ }\textbf {\bibinfo
  {volume} {6}},\ \bibinfo {pages} {3267--3275} (\bibinfo {year}
  {1994})}\BibitemShut {NoStop}%
\bibitem [{\citenamefont {Maruvada}\ and\ \citenamefont
  {Park}(1996)}]{Maruvada_1996a}%
  \BibitemOpen
  \bibfield  {author} {\bibinfo {author} {\bibfnamefont {S.~R.~K.}\
  \bibnamefont {Maruvada}}\ and\ \bibinfo {author} {\bibfnamefont {C.-W.}\
  \bibnamefont {Park}},\ }\bibfield  {title} {\enquote {\bibinfo {title}
  {Retarded motion of bubbles in {Hele--Shaw} cells},}\ }\href@noop {}
  {\bibfield  {journal} {\bibinfo  {journal} {Phys.~Fluids}\ }\textbf {\bibinfo
  {volume} {8}},\ \bibinfo {pages} {3229--3233} (\bibinfo {year}
  {1996})}\BibitemShut {NoStop}%
\bibitem [{\citenamefont {Afkhami}\ and\ \citenamefont
  {Renardy}(2013)}]{Afkhami_2013a}%
  \BibitemOpen
  \bibfield  {author} {\bibinfo {author} {\bibfnamefont {S.}~\bibnamefont
  {Afkhami}}\ and\ \bibinfo {author} {\bibfnamefont {Y.}~\bibnamefont
  {Renardy}},\ }\bibfield  {title} {\enquote {\bibinfo {title} {A
  volume-of-fluid formulation for the study of co-flowing fluids governed by
  the {Hele--Shaw} equations},}\ }\href@noop {} {\bibfield  {journal} {\bibinfo
   {journal} {Phys.~Fluids}\ }\textbf {\bibinfo {volume} {25}},\ \bibinfo
  {pages} {082001} (\bibinfo {year} {2013})}\BibitemShut {NoStop}%
\bibitem [{\citenamefont {Gallaire}\ \emph {et~al.}(2014)\citenamefont
  {Gallaire}, \citenamefont {Meliga}, \citenamefont {Laure},\ and\
  \citenamefont {Baroud}}]{Gallaire_2014a}%
  \BibitemOpen
  \bibfield  {author} {\bibinfo {author} {\bibfnamefont {F.}~\bibnamefont
  {Gallaire}}, \bibinfo {author} {\bibfnamefont {P.}~\bibnamefont {Meliga}},
  \bibinfo {author} {\bibfnamefont {P.}~\bibnamefont {Laure}}, \ and\ \bibinfo
  {author} {\bibfnamefont {C.~N.}\ \bibnamefont {Baroud}},\ }\bibfield  {title}
  {\enquote {\bibinfo {title} {Marangoni induced force on a drop in a {Hele
  Shaw} cell},}\ }\href@noop {} {\bibfield  {journal} {\bibinfo  {journal}
  {Phys.~Fluids}\ }\textbf {\bibinfo {volume} {26}},\ \bibinfo {pages} {062105}
  (\bibinfo {year} {2014})}\BibitemShut {NoStop}%
\bibitem [{\citenamefont {Huerre}\ \emph {et~al.}(2015)\citenamefont {Huerre},
  \citenamefont {Theodoly}, \citenamefont {Leshansky}, \citenamefont
  {Valignat}, \citenamefont {Cantat},\ and\ \citenamefont
  {Jullien}}]{Huerre_2015a}%
  \BibitemOpen
  \bibfield  {author} {\bibinfo {author} {\bibfnamefont {A.}~\bibnamefont
  {Huerre}}, \bibinfo {author} {\bibfnamefont {O.}~\bibnamefont {Theodoly}},
  \bibinfo {author} {\bibfnamefont {A.~M.}\ \bibnamefont {Leshansky}}, \bibinfo
  {author} {\bibfnamefont {M.-P.}\ \bibnamefont {Valignat}}, \bibinfo {author}
  {\bibfnamefont {I.}~\bibnamefont {Cantat}}, \ and\ \bibinfo {author}
  {\bibfnamefont {M.-C.}\ \bibnamefont {Jullien}},\ }\bibfield  {title}
  {\enquote {\bibinfo {title} {Droplets in microchannels: Dynamical properties
  of the lubrication film},}\ }\href@noop {} {\bibfield  {journal} {\bibinfo
  {journal} {Phys.~Rev.~Lett.}\ }\textbf {\bibinfo {volume} {115}},\ \bibinfo
  {pages} {064501} (\bibinfo {year} {2015})}\BibitemShut {NoStop}%
\bibitem [{\citenamefont {Afkhami}, \citenamefont {Leshansky},\ and\
  \citenamefont {Renardy}(2011)}]{Afkhami_2011a}%
  \BibitemOpen
  \bibfield  {author} {\bibinfo {author} {\bibfnamefont {S.}~\bibnamefont
  {Afkhami}}, \bibinfo {author} {\bibfnamefont {A.~M.}\ \bibnamefont
  {Leshansky}}, \ and\ \bibinfo {author} {\bibfnamefont {Y.}~\bibnamefont
  {Renardy}},\ }\bibfield  {title} {\enquote {\bibinfo {title} {Numerical
  investigation of elongated drops in a microfluidic {T-junction}},}\
  }\href@noop {} {\bibfield  {journal} {\bibinfo  {journal} {Phys.~Fluids}\
  }\textbf {\bibinfo {volume} {23}},\ \bibinfo {pages} {022002} (\bibinfo
  {year} {2011})}\BibitemShut {NoStop}%
\bibitem [{\citenamefont {Bedram}\ and\ \citenamefont
  {Moosavi}(2013)}]{Bedram_2013a}%
  \BibitemOpen
  \bibfield  {author} {\bibinfo {author} {\bibfnamefont {A.}~\bibnamefont
  {Bedram}}\ and\ \bibinfo {author} {\bibfnamefont {A.}~\bibnamefont
  {Moosavi}},\ }\bibfield  {title} {\enquote {\bibinfo {title} {Breakup of
  droplets in micro and nanofluidic {T-Junctions}},}\ }\href@noop {} {\bibfield
   {journal} {\bibinfo  {journal} {J. Appl. Fluid Mech.}\ }\textbf {\bibinfo
  {volume} {6}},\ \bibinfo {pages} {81--86} (\bibinfo {year}
  {2013})}\BibitemShut {NoStop}%
\bibitem [{\citenamefont {Hoang}\ \emph
  {et~al.}(2013{\natexlab{a}})\citenamefont {Hoang}, \citenamefont {{van
  Steijn}}, \citenamefont {Portela}, \citenamefont {Kreutzer},\ and\
  \citenamefont {Kleijn}}]{Hoang_2013a}%
  \BibitemOpen
  \bibfield  {author} {\bibinfo {author} {\bibfnamefont {D.~A.}\ \bibnamefont
  {Hoang}}, \bibinfo {author} {\bibfnamefont {V.}~\bibnamefont {{van Steijn}}},
  \bibinfo {author} {\bibfnamefont {L.~M.}\ \bibnamefont {Portela}}, \bibinfo
  {author} {\bibfnamefont {M.~T.}\ \bibnamefont {Kreutzer}}, \ and\ \bibinfo
  {author} {\bibfnamefont {C.~R.}\ \bibnamefont {Kleijn}},\ }\bibfield  {title}
  {\enquote {\bibinfo {title} {Benchmark numerical simulations of segmented
  two-phase flows in microchannels using the volume of fluid method},}\
  }\href@noop {} {\bibfield  {journal} {\bibinfo  {journal} {Comput.~Fluids}\
  }\textbf {\bibinfo {volume} {86}},\ \bibinfo {pages} {28--36} (\bibinfo
  {year} {2013}{\natexlab{a}})}\BibitemShut {NoStop}%
\bibitem [{\citenamefont {Hoang}\ \emph
  {et~al.}(2013{\natexlab{b}})\citenamefont {Hoang}, \citenamefont {Portela},
  \citenamefont {Kleijn}, \citenamefont {Kreutzer},\ and\ \citenamefont {{van
  Steijn}}}]{Hoang_2013b}%
  \BibitemOpen
  \bibfield  {author} {\bibinfo {author} {\bibfnamefont {D.~A.}\ \bibnamefont
  {Hoang}}, \bibinfo {author} {\bibfnamefont {L.~M.}\ \bibnamefont {Portela}},
  \bibinfo {author} {\bibfnamefont {C.~R.}\ \bibnamefont {Kleijn}}, \bibinfo
  {author} {\bibfnamefont {M.~T.}\ \bibnamefont {Kreutzer}}, \ and\ \bibinfo
  {author} {\bibfnamefont {V.}~\bibnamefont {{van Steijn}}},\ }\bibfield
  {title} {\enquote {\bibinfo {title} {Dynamics of droplet breakup in a
  {T-junction}},}\ }\href@noop {} {\bibfield  {journal} {\bibinfo  {journal}
  {J. Fluid Mech.}\ }\textbf {\bibinfo {volume} {717}},\ \bibinfo {pages} {R4}
  (\bibinfo {year} {2013}{\natexlab{b}})}\BibitemShut {NoStop}%
\bibitem [{\citenamefont {Zhu}\ and\ \citenamefont
  {Gallaire}(2016)}]{Zhu_2016a}%
  \BibitemOpen
  \bibfield  {author} {\bibinfo {author} {\bibfnamefont {L.}~\bibnamefont
  {Zhu}}\ and\ \bibinfo {author} {\bibfnamefont {F.}~\bibnamefont {Gallaire}},\
  }\href@noop {} {\enquote {\bibinfo {title} {A pancake droplet translating in
  a {Hele--Shaw} cell: lubrication film and flow field},}\ } (\bibinfo {year}
  {2016}),\ \bibinfo {note} {arXiv:1601.08157}\BibitemShut {NoStop}%
\bibitem [{\citenamefont {Popinet}(2003)}]{Popinet_2003a}%
  \BibitemOpen
  \bibfield  {author} {\bibinfo {author} {\bibfnamefont {S.}~\bibnamefont
  {Popinet}},\ }\bibfield  {title} {\enquote {\bibinfo {title} {Gerris: a
  tree-based adaptive solver for the incompressible euler equations in complex
  geometries},}\ }\href@noop {} {\bibfield  {journal} {\bibinfo  {journal}
  {J.~Comput.~Phys.}\ }\textbf {\bibinfo {volume} {190}},\ \bibinfo {pages}
  {572--600} (\bibinfo {year} {2003})}\BibitemShut {NoStop}%
\bibitem [{\citenamefont {Popinet}(2009)}]{Popinet_2009a}%
  \BibitemOpen
  \bibfield  {author} {\bibinfo {author} {\bibfnamefont {S.}~\bibnamefont
  {Popinet}},\ }\bibfield  {title} {\enquote {\bibinfo {title} {An accurate
  adaptive solver for surface-tension-driven interfacial flows},}\ }\href@noop
  {} {\bibfield  {journal} {\bibinfo  {journal} {J.~Comput.~Phys.}\ }\textbf
  {\bibinfo {volume} {228}},\ \bibinfo {pages} {5838--5866} (\bibinfo {year}
  {2009})}\BibitemShut {NoStop}%
\bibitem [{\citenamefont {Scardovelli}\ and\ \citenamefont
  {Zaleski}(1999)}]{Scardovelli_1999a}%
  \BibitemOpen
  \bibfield  {author} {\bibinfo {author} {\bibfnamefont {R.}~\bibnamefont
  {Scardovelli}}\ and\ \bibinfo {author} {\bibfnamefont {S.}~\bibnamefont
  {Zaleski}},\ }\bibfield  {title} {\enquote {\bibinfo {title} {Direct
  numerical simulation of free-surface and interfacial flow},}\ }\href@noop {}
  {\bibfield  {journal} {\bibinfo  {journal} {Annu.~Rev.~Fluid Mech.}\ }\textbf
  {\bibinfo {volume} {31}},\ \bibinfo {pages} {567--603} (\bibinfo {year}
  {1999})}\BibitemShut {NoStop}%
\bibitem [{\citenamefont {Tryggvason}, \citenamefont {Scardovelli},\ and\
  \citenamefont {Zaleski}(2011)}]{Tryggvason_2011a}%
  \BibitemOpen
  \bibfield  {author} {\bibinfo {author} {\bibfnamefont {G.}~\bibnamefont
  {Tryggvason}}, \bibinfo {author} {\bibfnamefont {R.}~\bibnamefont
  {Scardovelli}}, \ and\ \bibinfo {author} {\bibfnamefont {S.}~\bibnamefont
  {Zaleski}},\ }\href@noop {} {\emph {\bibinfo {title} {Direct numerical
  simulations of gas-liquid multiphase flows}}}\ (\bibinfo  {publisher}
  {Cambridge University Press},\ \bibinfo {year} {2011})\BibitemShut {NoStop}%
\bibitem [{\citenamefont {Francois}\ \emph {et~al.}(2006)\citenamefont
  {Francois}, \citenamefont {Cummins}, \citenamefont {Dendy}, \citenamefont
  {Kothe}, \citenamefont {Sicilian},\ and\ \citenamefont
  {Williams}}]{Francois_2006a}%
  \BibitemOpen
  \bibfield  {author} {\bibinfo {author} {\bibfnamefont {M.~M.}\ \bibnamefont
  {Francois}}, \bibinfo {author} {\bibfnamefont {S.~J.}\ \bibnamefont
  {Cummins}}, \bibinfo {author} {\bibfnamefont {E.~D.}\ \bibnamefont {Dendy}},
  \bibinfo {author} {\bibfnamefont {D.~B.}\ \bibnamefont {Kothe}}, \bibinfo
  {author} {\bibfnamefont {J.~M.}\ \bibnamefont {Sicilian}}, \ and\ \bibinfo
  {author} {\bibfnamefont {M.~W.}\ \bibnamefont {Williams}},\ }\bibfield
  {title} {\enquote {\bibinfo {title} {A balanced-force algorithm for
  continuous and sharp interfacial surface tension models within a volume
  tracking framework},}\ }\href@noop {} {\bibfield  {journal} {\bibinfo
  {journal} {J.~Comput.~Phys.}\ }\textbf {\bibinfo {volume} {213}},\ \bibinfo
  {pages} {141--173} (\bibinfo {year} {2006})}\BibitemShut {NoStop}%
\bibitem [{\citenamefont {Brackbill}, \citenamefont {Kothe},\ and\
  \citenamefont {Zemach}(1992)}]{Brackbill_1992a}%
  \BibitemOpen
  \bibfield  {author} {\bibinfo {author} {\bibfnamefont {J.~U.}\ \bibnamefont
  {Brackbill}}, \bibinfo {author} {\bibfnamefont {D.~B.}\ \bibnamefont
  {Kothe}}, \ and\ \bibinfo {author} {\bibfnamefont {C.}~\bibnamefont
  {Zemach}},\ }\bibfield  {title} {\enquote {\bibinfo {title} {A continuum
  method for modeling surface tension},}\ }\href@noop {} {\bibfield  {journal}
  {\bibinfo  {journal} {J.~Comput.~Phys.}\ }\textbf {\bibinfo {volume} {100}},\
  \bibinfo {pages} {335--354} (\bibinfo {year} {1992})}\BibitemShut {NoStop}%
\bibitem [{\citenamefont {Landau}\ and\ \citenamefont
  {Levich}(1942)}]{Landau_1942a}%
  \BibitemOpen
  \bibfield  {author} {\bibinfo {author} {\bibfnamefont {L.}~\bibnamefont
  {Landau}}\ and\ \bibinfo {author} {\bibfnamefont {B.}~\bibnamefont
  {Levich}},\ }\bibfield  {title} {\enquote {\bibinfo {title} {Dragging of a
  liquid by a moving plate},}\ }\href@noop {} {\bibfield  {journal} {\bibinfo
  {journal} {Acta Physicochim. (URSS)}\ }\textbf {\bibinfo {volume} {17}},\
  \bibinfo {pages} {42--54} (\bibinfo {year} {1942})}\BibitemShut {NoStop}%
\bibitem [{\citenamefont {Galusinski}\ and\ \citenamefont
  {Vigneaux}(2008)}]{Galusinski_2008a}%
  \BibitemOpen
  \bibfield  {author} {\bibinfo {author} {\bibfnamefont {C.}~\bibnamefont
  {Galusinski}}\ and\ \bibinfo {author} {\bibfnamefont {P.}~\bibnamefont
  {Vigneaux}},\ }\bibfield  {title} {\enquote {\bibinfo {title} {On stability
  condition for bifluid flows with surface tension: Application to
  microfluidics},}\ }\href@noop {} {\bibfield  {journal} {\bibinfo  {journal}
  {J.~Comput.~Phys.}\ }\textbf {\bibinfo {volume} {227}},\ \bibinfo {pages}
  {6140--6164} (\bibinfo {year} {2008})}\BibitemShut {NoStop}%
\bibitem [{\citenamefont {Baroud}, \citenamefont {Gallaire},\ and\
  \citenamefont {Dangla}(2010)}]{Baroud_2010a}%
  \BibitemOpen
  \bibfield  {author} {\bibinfo {author} {\bibfnamefont {C.~N.}\ \bibnamefont
  {Baroud}}, \bibinfo {author} {\bibfnamefont {F.}~\bibnamefont {Gallaire}}, \
  and\ \bibinfo {author} {\bibfnamefont {R.}~\bibnamefont {Dangla}},\
  }\bibfield  {title} {\enquote {\bibinfo {title} {Dynamics of microfluidic
  droplets},}\ }\href@noop {} {\bibfield  {journal} {\bibinfo  {journal} {Lab
  Chip.}\ }\textbf {\bibinfo {volume} {10}},\ \bibinfo {pages} {2032--2045}
  (\bibinfo {year} {2010})}\BibitemShut {NoStop}%
\end{thebibliography}
%

\end{document}